\documentclass[fleqn,usenatbib]{mnras}

\usepackage{newtxtext,newtxmath}
\usepackage{url}
\usepackage{color}
\usepackage{tabulary}
\usepackage{multirow}
\usepackage{amsfonts}
\usepackage{array}
\usepackage{ulem}
\usepackage[T1]{fontenc}
\usepackage{ae,aecompl}

\usepackage{graphicx}	
\usepackage{amsmath}	
\usepackage{amssymb}	

\graphicspath{ {pictures/} }
\newcommand{\beq}{\begin{equation}}
\newcommand{\eeq}{\end{equation}}
\newcommand{\simlt}{\mathrel{\hbox{\rlap{\hbox{\lower4pt\hbox{$\sim$}}}\hbox{$<$}}}}
\newcommand{\simgt}{\mathrel{\hbox{\rlap{\hbox{\lower4pt\hbox{$\sim$}}}\hbox{$>$}}}}

\newcommand{\s}{\;\mathrm{s}}

\newcommand{\Msol}{\;\mathrm{M}_{\odot}}
\newcommand{\cm}{\;\mathrm{cm}}
\newcommand{\km}{\;\mathrm{km}}
\newcommand{\cmcube}{\;\mathrm{cm}^{-3}}
\newcommand{\AU}{\;\mathrm{AU}}
\newcommand{\pc}{\;\mathrm{pc}}
\newcommand{\yr}{\;\mathrm{yr}}

\def\apjl{ApJL}
\def\apj{ApJ}
\def\mnras{M.N.R.A.S.}
\def\aap{A\&A}
\def\nat{Nat.}
\def\aplett{ApL}

\def\apjs{ApJ Supp.}
\def\aj{AJ}

\def\apss{A\&Sp.Sc.}

\def\aapr{A\&A Rev.}

\title[binary-binary encounters]{Numerical study of $N=4$ binary-binary scatterings in a background potential}

\author[T. Ryu et al.]{
Taeho Ryu$^{1}$\thanks{email: taeho.ryu@stonybrook.edu}, Nathan W. C. Leigh$^{2}$, Rosalba Perna$^{1}$
\\
$^{1}$Department of Physics and Astronomy, Stony Brook University, Stony Brook, NY 11794-3800, USA\\
$^{2}$Department of Astrophysics, American Museum of Natural History, Central Park West and 79th Street, New York, NY 10024}

\date{Accepted XXX. Received YYY; in original form ZZZ}

\pubyear{2016}

\begin{document}
\label{firstpage}
\pagerange{\pageref{firstpage}--\pageref{lastpage}}
\maketitle

\begin{abstract}
	We perform a large suite of $N=4$ numerical scattering
	experiments between two identical binaries consisting of
	identical point particles in a (continuous) background
	potential.  For investigative purposes, we assume a uniform
	(natal or star-forming) gas medium. We explore a range of
	constant gas densities, from $n=10\cmcube$ to $10^{5}\cmcube$.
	These densities are relevant for various astrophysical
	environments, including star-forming molecular clouds and
	denser, fragmented cores within these clouds. Our primary goal
	is to characterize the effects of the background potential on
	the subsequent stellar dynamics.  We consider the outcome
	probabilities and the properties of any binaries formed during
	the binary-binary encounters, such as the distributions of
	binary binding energies and eccentricities. We also present
	the final velocity distributions of the ejected single stars.
	
	The background potential has two important effects on the stellar dynamics: 
	1) The potential acts to 
	reset the zero-point of the total system energy, which affects the types and 
	properties of the products of the encounter; 2) 
	For higher $n$ and weakly bound systems (i.e., large semimajor axes), 
	the stellar dynamics are significantly affected when stars become trapped in the 
	potential, oscillating around the system centre of 
	mass (CM). This, in turn, increases 
	the number of scattering events between stars (single, binary or triple) near the CM 
	and makes it harder for single stars to escape to infinity.  This ultimately 
	leads to the preferential ionization of triples and wide binaries and the survival of 
	compact binaries, with the single stars escaping at very high ejection velocities. 
\end{abstract}




\section{introduction}

\label{sec: introduction}

The $N$-body problem has been a topic intensively investigated by 
a number of studies using numerical simulations. There have been 
enormous achievements in studies of 3-body 
\citep[e.g][]{Heggie+93,Mikkola1994,Heggie+96} and 
4-body scattering problems \citep[e.g][]{Mikkola1983,Mikkola1984,Leigh+16} 
along with increasing computational capability, which render us a 
more extensive and deeper understanding of the dynamical phenomena 
occurring over the course of the evolution of stellar clusters. 
However, our understanding of these processes remains incomplete.

Binary-binary encounters dominate over binary-single encounters in star 
clusters with binary fractions $\gtrsim$ 10\% \citep{Sigurdsson93,Leigh2011}.  
Hence, binary-binary interactions are particularly important in dense stellar 
environments with high binary fractions.  This is the case ($\gtrsim 30-50\%$) in 
open clusters \citep[e.g][]{Halbwachs2003,Sana+08,Sana+09,Sollima+2010,Sana+2011,LeighGeller2013} 
and the solar neighborhood \citep[e.g][]{Abt1976,Duquennoy1991,Reid1997,Cool2002}. 
It follows that binary stars can play an important role in the evolution of star clusters 
with high binary fractions.  The binary frequencies in the cores of globular 
clusters are relatively low ($\lesssim10\%$, \citealt[e.g][]{Rubenstein1997,Albrow+2001,Ji2015}), 
such that binary-single interactions can dominate over binary-binary interactions \citep{Sigurdsson93,Leigh2011}.  
Regardless, it is generally accepted that close encounters between binaries occur commonly 
in globular clusters \citep{Hills1976,Hut83}. The binary fractions in the cores of clusters 
tend to be elevated relative to the outskirts due to mass segregation induced by 
two-body relaxation.  Binary-binary encounters should produce triple star systems commonly \citep[e.g.][]{Leigh+16}.  
Triples can also form via dissipative tidal captures \citep[e.g][]{Bailyn89} or scatterings involving 
pre-existing triples \citep[e.g.][]{Leigh2011,LeighGeller2012,LeighGeller2013,leigh15,Antognini2016}.

Dynamical scatterings between binaries provide a formation channel 
for compact binary systems. 
In general, hard binaries are hardened further during binary-binary encounters 
via the slingshot mechanism, while soft binaries are ionized \citep{Heggie1975,Hills1976}.  
In this context, of particular interest are the three gravitational wave events recently reported by aLIGO, i.e. 
GW150914, GW151226 and LVT151012 \citep{Abbott+2016all}. These are believed to 
have originated from the coalescence 
of two massive black holes in a binary system. Many authors have pointed out that the 
formation of compact binary black holes via stellar scatterings in dense star clusters is a feasible 
mechanism for the formation of these binary black holes, and that the measured BH masses 
are consistent with this formation channel 
\citep[and references therein]{PortegiesZwart2000,Abbott+2016GW150914}.
At high redshifts ($z\sim15-20$), compact (and wider) black hole binaries, such as 
high mass X-ray binaries, consisting of the first remnants from Population III stars 
could form via binary-binary interactions 
in star-forming clouds \citep{Ryu+2016}.  Once formed, these black hole binaries could 
evolve into even more compact binaries 
(depending on the initial binary orbital parameters), possibly contributing to 
future GW detections by Advanced LIGO \citep[e.g.][]{Belczynski+2016}.  
In addition to hardening binaries, physical collisions (and mergers) could occur when two 
binaries interact directly \citep{Hills1976,Fregeau+2004,LeighGeller2012,leigh15}.  This has been 
discussed as a viable mechanism for the formation of blue stragglers, 
observed in both open and globular clusters \citep{Leonard1989,Davies+2004,Leigh+2007,Knigge2009}.

To date, most authors studying small-number scattering interactions have focused on 
isolated stellar dynamics (analytically and numerically).  However, observational 
evidence suggests that all dense clusters experience a phase 
where gas and stars co-exist \citep{Brown1975,Samson1975}.
Here, stars and binaries undergo direct interactions in the presence of a 
background gas medium. This phase may be a natural consequence of the formation 
of dense clusters. Briefly, stars form out of cool, dense gas.  
Eventually, the gas is removed from the cluster, possibly due to stellar winds 
and/or supernovae explosions. This reduces the binding energy of the cluster, 
liberating stars from the outskirts in the process. 
The existence of multiple stellar populations in globular clusters 
\citep[and references therein]{Gratton+2012} 
suggests that this could have occurred more than once over the cluster life time \citep{Lin2007}.
Therefore, unveiling the stellar dynamics in a background gas medium 
may be essentially the first step 
in connecting the gap between the gas-dominated star formation regime and 
the gas-free stellar dynamics regime. 

The goal of this paper is to characterize the effects of a continuous background potential 
on the time evolution of binary-binary interactions.  
To this end, we present the results of a large suite of numerical $N$-body 
simulations of binary-binary scatterings in a background potential. For simplicity, we model 
the background potential as an uniform 
(natal or star-forming) gas medium.  \textit{We consider 
only the gravitational effects of the background potential,} and ignore any gas damping 
effects (dynamical friction, accretion, etc.).  Hence, our choice for the gas density directly 
determines the total mass of the potential for a given outer boundary, or, equivalently, it 
determines the location of the outer boundary for a given total gas mass.  We consider a range of (discrete) 
constant gas densities, from $n=10\cmcube$ to $10^{5}\cmcube$.  These densities 
are typical for molecular clouds (i.e. regions of star formation) 

We primarily focus on the outcome probabilities and the 
properties of any binaries formed during the encounters. 
We find that one of the important roles of the 
background potential (for any non-zero number density) is to reset the zero-point of 
the total system energy.  This in turn affects the types of objects formed during the 
encounters in addition to the outcome 
probabilities. In particular, for higher $n$ and less tightly bound 
systems (i.e., larger semimajor axes), stars often become 
trapped in the background potential, oscillating with a short period around the system center of mass. 
This results in fewer triples and wider binaries and the preferential survival of 
compact binaries, since subsequent scatterings continue until two single stars escape 
at very high ejection velocities. Note that we only consider the gravitational influence of 
the background potential and do not consider 
gas dynamical friction, gas damping or gas accretion. 
We emphasize that our results are general to any continuous background potential (e.g., dark matter), 
and are not just limited to a constant gas medium. Hence, we shall use throughout 
this paper ``continuous background potential"  instead of gravitational 
gas potential.

The paper is organized as follows. We start in \S\ref{sec:method} by describing 
the implementation of the background potential in our scattering code, including 
the code termination criteria and our choices for the initial conditions. 
We present our results in \S\ref{sec:results}, including the outcome probabilities 
and the properties of any binaries formed during the interactions.  
Finally, we summarize the effects of the background potential on the stellar 
dynamics in our simulations, and discuss the implications of our work for real 
astrophysical environments in \S\ref{sec:discussionandsummary}.

\section{Method}
\label{sec:method}
In this section, we provide descriptions of the numerical scattering simulations 
of binary-binary encounters, the background gas medium and the 
termination criteria adopted for the simulations. 

\subsection{Overview of the scattering experiment}
\label{sec:overview}
We investigate the products of binary-binary encounters in a continuous 
background potential. The equations of motion of four stars are solved with the 
4th - order \& 5 - stage Runge-Kutta-Fehlberg method \citep{Fehlberg} using 
adaptive time steps with the error control tolerance
\footnote{In the Runge-Kutta-Fehlberg
		method, the error can be controlled by using the higher-order
		embedded method. The error, defined as a difference between two
		solutions from 4th - order and 5th - order calculations, is
		estimated at each time step, and the following time step is
		automatically determined to give the error less than a given error
		control tolerance.} of $10^{-12}$.
	 The numerical method is a very precise and stable integration 
method among the large class of Runge-Kutta schemes, particularly by adopting 
the Butcher tableau for Fehlberg's 4(5) method.
However, a small error control tolerance can lead to a small time step size. 
	So in order to maintain an acceptable computation speed throughout the simulations, 
	we additionally set a minimum value for the time step, 
	$10^{-7}\times \tau_{\rm dyn, \,min}$, where $\tau_{\rm dyn, \,min}$ 
	is the smallest value of the dynamical time between
	any two stars in the simulation at a given time step. 
	With this, the actual numerical errors could be higher than what was 
	basically set by the error control tolerance, but we can achieve both of an acceptable computation speed 
	and still small numerical errors in the total energy. 
	Given the initial total energy of the system $E(t=0)$,
	the numerical error in the total energy of the system 
	($|[E(t)-E(t=0)]/E(t=0)|$ where $E(t)$ is defined in the equation \ref{eq5})
	never exceeds $10^{-4}-10^{-6}$ in all simulations.  

We perform $10^{3}$ scattering experiments of two identical 
binaries with the same initial 
semimajor axis for every given set of simulations. The number of 
simulations for each set is 
sufficiently large to ensure that all data statistically converge 
(the overall Poisson uncertainties $\lesssim 3-4\%$).
 In total, we consider three discrete 
values of the semimajor axis ($a_{0}=1\AU, 10\AU$ and $100\AU$). We first generate 
two identical binaries with initially circular orbits (the eccentricities are zero, or $e=0$). 
Each binary consists of two point particles, each with a mass of $1\Msol$. Then we 
give an initial separation and relative velocity between the centre of mass of the two 
binaries, such that they collide head-on at the origin (i.e., with impact parameter $b=0$) 
with nearly zero total energy (more precisely, slightly negative to positive depending 
on the number density of the gas medium, but we defer a more detailed explanation 
of this to section \ref{sec:gasmedium}). However, the mutual inclinations between 
the binary orbital planes, as well as their initial phases, are randomly chosen.

In order to choose the value of the initial relative velocity $v_{\rm rel}$  
	between the centres of mass of the two binaries, 
we define the critical velocity 
$v_{\rm cri}$ as the relative velocity between the centres of mass of the two binaries 
at infinity for which the total encounter energy 
is zero. In this paper, $v_{\rm rel}$ will be presented in units of $v_{\rm cri}$, unless 
	otherwise stated. More explicitly, in the case of binary-binary encounters, the total 
initial energy $E_{i}$ at infinity is written as follows:

\begin{align}
\label{eq:1}
E_{i}&=\frac{1}{2}\mu v_{\rm rel}^{2} - \frac{Gm_{11}m_{12}}{2a_{1}} - \frac{G m_{21}m_{22}}{2a_{2}}.
\end{align}
In this equation, the first subscript for the particles mass $m$, along with the binary semimajor axis 
$a$ subscript, corresponds to each binary, whereas the second subscript for $m$ 
corresponds to each component of the binary. For example, the masses of the two stars 
in the $1_{\rm st}$ binary with semimajor axis $a_{1}$ are $m_{11}$ and $m_{12}$. 
Here, $\mu$ in the first term on the right hand side is the reduced mass of the two binaries, i.e., 
	$\mu=(m_{11}+m_{12})(m_{21}+m_{22})/(m_{11}+m_{12}+m_{21}+m_{22})=1\Msol$. The second 
and third terms are the orbital energies of the two binaries. 
Note that the gravitational potential between the two binaries is zero at infinity. 
Therefore, the critical velocity at $E_{i}=0$ is,

\begin{align}
\label{eq:2}
v_{\rm cri}&=v_{\rm rel}(E_{i}=0)=\sqrt{\frac{2}{\mu}\Big(\frac{Gm_{11}m_{12}}{2a_{1}}+\frac{Gm_{21}m_{22}}{2a_{2}}\Big)}\,.
\end{align}
In this study, we take $v_{\rm rel}=1$ to explore the zero-energy limit. 

In the presence of a background gas medium, the condition $v_{\rm rel}=1$ does 
not necessarily imply that the total encounter energy is exactly zero. Instead, we 
achieve the zero-energy limit by adjusting the initial separations between the binary 
centers of mass. In order to explain how we determine the initial distance between 
the two binaries, we first need to introduce the background potential adopted in 
this study, corresponding to a constant density gas medium.

\subsection{Background gravitational potential}
\label{sec:gasmedium}

In this section, we describe the background gravitational potential adopted in our experiments, the choice 
of initial conditions, and the resulting motions of the stars. 

We assume a uniform (i.e. constant) density $\rho$ ($n$ is
	the gas number density) for this potential, with a limiting outer
	boundary $r_{\rm bg}$ set by our chosen total background mass
	$M_{\rm gas}$ such that
	\begin{align}
		\rho&=
		\begin{cases}
			n m_{\rm H} \hfill \hspace{0.4in} r \leq r_{\rm bg}\,;\\
			0 \hfill r>r_{\rm bg}\,,
		\end{cases}
		\label{eq2}
	\end{align}
	where $m_{\rm H}$ is the mass of a hydrogen atom. For simplicity, we take a mean molecular weight of unity.
	The mass of gas enclosed in a spherical volume of radius $r$ can be written as
	\begin{align}
		M_{\rm en,gas}(r)&=
		\begin{cases}
			\frac{4\pi}{3}\rho r^{3} \hfill \hspace{0.8in} r \leq r_{\rm bg}\,;\\
			\frac{4\pi}{3}\rho r_{\rm bg}^{3}=M_{\rm gas} \hfill r>r_{\rm bg}\,.
		\end{cases}
		\label{eq:engasmass}
	\end{align}

 We consider a range of densities, motivated 
 by the values typical of gaseous astrophysical environments, most notably giant 
 molecular clouds and star-forming regions.  In this case, 
 the gravitational force imparted by a uniform gas medium of constant density 
 on a given star particle at $r$ follows the analytic formula: 
 	\begin{align}
 		\label{eq3}
 		\textbf{f}_{\rm bg}(r)&=-\frac{Gm M_{\rm en,gas}(r)}{r^{3}}\textbf{r}\nonumber\\ 
 		&=
 		\begin{cases}
 			-\frac{4}{3}\pi G m\rho \;\textbf{r} \hfill \hspace{0.3in} r \leq r_{\rm bg}\,;\\
 			-\frac{4}{3}\pi G m\rho \Big(\frac{r_{\rm bg}}{r}\Big)^{3}\textbf{r} \hfill \hspace{0.3in} r > r_{\rm bg}\,,\\
 		\end{cases}
 	\end{align}
 	where $m$ is the mass of the star and $\textbf{r}$ is the vector pointing from the system CM to the star. Accordingly, the background potential has the following form:
 	\begin{align}
 		\label{eq4}
 		V_{\rm bg}(r)&=
 		\begin{cases}
 			\frac{2}{3}\pi G m\rho (r^{2}-3r_{\rm bg}^{2})\hfill \hspace{0.3in} r \leq r_{\rm bg}\,;\\
 			-\frac{GmM_{\rm gas}}{r}=-\frac{4}{3}\pi G m\rho \frac{r_{\rm bg}^{3}}{r} \hfill \hspace{0.3in} r > r_{\rm bg}\,.\\
 		\end{cases}
 	\end{align}
 	And the total energy $E(t)$ of four stars in the system at $t$, including the contribution from $V_{\rm bg}(r)$, 
 	can be written as,
 	\begin{equation}
 		\label{eq5}
 		E(t)=\sum_{i=1}^{4}\frac{1}{2}m_{i}v_{i}^{2}-\sum_{\substack{i,j=1\\(i>j)}}^{4}
 		\frac{Gm_{i}m_{j}}{|\textbf{r}_{i}-\textbf{r}_{j}|}+\sum_{i=1}^{4} V_{\rm bg}(r_{i})\,,
 	\end{equation}
 	where $m_{i}$ is the mass of each star and $v_{i}$ is the velocity of each star with respect to the system CM.

 Note that we consider only the gravitational force imparted by the background potential, and 
 ignore any dissipative effects caused by the motions of macroscopic objects in a dense gaseous 
 medium. 
 The background potential is assumed to remain static in time.  There is also no 
 stellar mass growth via gas accretion or dissipative effects due to the gas. 
 
 In general, the background potential provides 
 additional gravitational forces that accelerate the stars initially, increasing the relative velocity at 
 impact. When the stars approach the system CM at the moment of impact, the influence of the 
 background potential becomes (instantaneously) small. On the other hand, when the star particles 
 are moving away from each other after scattering events or ejections, the potential may play an 
 important role as an inward restoring force.  This ultimately prevents or delays dissociation, 
 and increases the probability of subsequent encounters (depending on the gas density). 
 
In order to determine the initial separations $r_{12}$
 	between the centres of mass of the two binaries (located at initial
 	distances $r_{1}$ and $r_{2}$ from the system CM), we use the
 	monopole approximation for the binaries (i.e. a point particle
 with a mass of $2\Msol$) in equation \ref{eq5} 
 	to evaluate both of the gravitational potential energies 
 	(between the two binaries and from the background medium)
 	\footnote{For a given semimajor axis $a$ for a binary, the relative errors of the
 		gravitational potential energies due to the monopole
 		approximation $(\Delta V/V=|(V_{\rm with~monopole}-V_{\rm
 			without~monopole})/V_{\rm without~monopole}|)$ are on the order of $\Big(a/r_{12}\Big)^{2}$.
 		This approximation has a negligible effect
 		 on the results since in our simulations the initial 
 		separations are sufficiently larger than the semimajor axes of 
 		the binaries (i.e., $\Delta V /V\lesssim 10^{-6}$).}. Under the presence
 	of a background potential, this approximation allows us to derive
 	$r_{12}$ by solving one simple equation for a given total
 	energy. More explicitly, 
 	the initial total energy for the two binaries in the background potential
becomes:
 \begin{align}
 	\label{eq:7}
	E_{0}&=\frac{1}{2}\mu v_{\rm rel}^{2} - \frac{Gm_{11}m_{12}}{2a_{1}} - \frac{G m_{21}m_{22}}{2a_{2}}\nonumber\\
&
+\frac{2}{3}\pi G m_{1}\rho r_{1}^{2}+\frac{2}{3}\pi G m_{2}\rho r_{2}^{2}-\frac{G m_{1}m_{2}}{r_{12}}\nonumber\\
&=\frac{1}{2}\mu v_{\rm rel}^{2}-2\frac{Gm_{11}m_{12}}{2a_{1}}+\frac{4}{3}\pi G m_{1}\rho \left(\frac{r_{12}}{2}\right)^{2}-\frac{G m_{1}m_{2}}{r_{12}}\,.
 \end{align}
 In the second equality, we simplified the equation using the fact that the two binaries are 
 identical and the CM of the two binaries coincides with the origin initially, i.e. $r_{1}=r_{2}=r_{12}/2$. 
 Here, the gravitational potential energy between the two binaries (the last term) should not be ignored because the two binaries 
 are initially separated by a finite distance in our simulations. Since we take $v_{\rm rel}=1$, the first 
 two terms on the right hand-side of the second equality (corresponding to $E_{i}$ in equation \ref{eq:1}) always cancel out, regardless of $a_{1}$ (and $a_{2}=a_1$). 
 Hence, in the limit of $E_0\simeq0$, $r_{12}$ is determined 
 by the requirement that the last two terms cancel each other. Once $r_{12}$ and the relative position of each star with respect to the center of mass of each binary are decided, so are the position and velocity vectors of the four stars.

In our experiments, we fix $r_{12}$ such that $E_{0}\simeq0$ (but slightly positive) when $n=10^{2}\cmcube$. 
In order to explore the effects of the background potential, we run suites of 
simulations with a range of densities, i.e., $n=10, 10^{2}, 10^{3}$ and 
$10^{5}\cmcube$ (default set). 
For this default set, we take $M_{\rm gas}=10^{10}\Msol$ so that the width of the potential 
	($r_{\rm bg}$) is large enough for all particles to remain within 
	the spherical volume defined by this critical radius throughout the duration of the interaction 
	($r_{\rm bg}$ for the chosen total gas mass is depicted in Figure \ref{fig:a_n_relation}).
For comparison, we run additional simulations with even 
higher number densities and larger semimajor axes. As we will show, the background potential 
resets the zero-point of the total system energy. In particular, the stellar 
dynamics are significantly affected by the background potential when the stars become 
trapped in the potential, oscillating with a short period around the system CM.  
This leads to repeated scattering events that would not otherwise occur without the influence of 
the background potential.

\begin{figure}
	\centering
	\includegraphics[width=8.9cm]{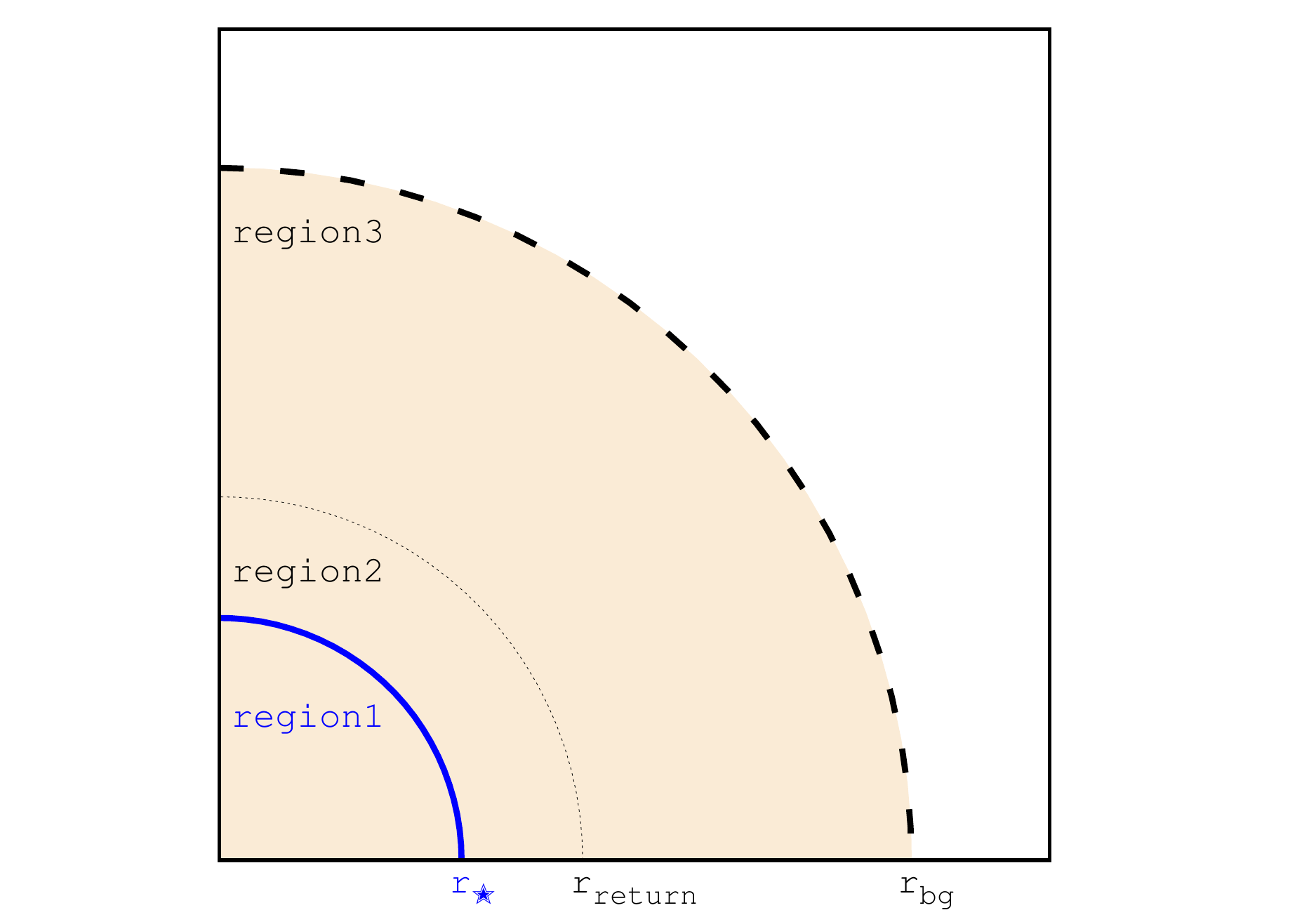}
	
	\caption{A schematic diagram projected on to the 2-D $x - y$ plane showing the three regions 
		defined by the three characteristic distances, namely $r_{\star}$, $r_{\rm return}$ and $r_{\rm bg}$. 
		In the inner-most region (blue solid circle) bounded by $r_{\star}$, the enclosed 
		gas mass $M_{\rm en,gas}$ is smaller than the total stellar mass $M_{\star}$. 
		$r_{\rm return}$ (black dotted line) indicates the distance at which the 
		stars return towards the system CM, and $r_{\rm bg}$ corresponds to the outer boundary of the 
		gas cloud or potential well.}
	\label{fig:configuration}
\end{figure}

\subsection{Modified termination criteria}
\label{terminaltioncriteria}

Our primary interest in this paper is in the outcome probabilities for binary-binary 
scatterings.  Hence, it is critical to properly define when the interactions between 
star particles are complete.  The simulations terminate when no further scatterings
 will occur. To begin, we use the same criteria as described in \citet{Fregeau+2004}. 
These criteria consist of four conditions: 1) All single stars and bound hierarchies 
in a system are moving away from one another. 2) The total energy of the single stars
 and bound hierarchies are positive (without including the orbital energies of any 
 bound hierarchies). 3) Hierarchies such as triples are dynamically stable. Finally, 
 4) the relative force between stars in a bound system should be larger at apocenter
  than the tidal force from all other stars (i.e., bound hierarchies as a whole, if 
  formed) by a certain fraction, the so called tidal tolerance parameter $\delta$. 
  The relative force $F_{\rm rel}$ and the tidal force $F_{\rm tid, \star}$ on the 
  $i_{\rm th}$ binary at apocenter are: 

\begin{align}
\label{eq:rel_force}
F_{\rm rel}&=\frac{Gm_{i1}m_{i2}}{[a(1+e)]^{2}}\,,\\
\label{eq:tid_force}
F_{\rm tid,\star}&=\sum_{j}\frac{2 G m_{i}m_{j}}{r_{ij}^{3}}a(1+e)\,.
\end{align}
Hence, the condition for the tidal tolerance parameter can be written:
\begin{align}
\label{eq:tidaltoleranceparameter}
\frac{F_{\rm tid,\star}}{F_{\rm rel}}<\delta.
\end{align}
We take $\delta=10^{-9}$, which is a value small enough to allow us to precisely classify the outcomes and explore the 
	effects of the background potential at large distances. For more details, see \citet{Fregeau+2004}.
Additional termination criteria beyond those implemented in \citet{Fregeau+2004} are needed 
in our case.  That is, in order to implement physically motivated termination criteria appropriate 
for scatterings occurring in the presence of a background potential, we must include additional 
criteria due to the tidal force and the relative force from the \textit{gas}. We calculate them at 
apocenter as follows:
	\begin{align}
	F_{\rm tid,bg}&=\frac{4\pi\rho}{3} G m_{i}[a(1+e)]\,,\\
	F_{\rm rel}&=\frac{Gm_{i}m_{j}}{[a(1+e)]^{2}}\Big|1-\Big(\frac{\frac{4\pi\rho}{3}[a(1+e)]^{3}}{m_{j}}\Big)\Big(\frac{r_{i}}{a(1+e)}\Big)\Big|\,.
\end{align}
We only consider the minimum relative force for a more strict termination criteria. We then 
impose two last additional termination criteria, namely a correction to the previous calculation 
of the tidal tolerance parameter and an additional escape criterion:

5) $\frac{F_{\rm tid,\star}+F_{\rm tid,bg}}{F_{\rm rel}}<\delta$\,,

6) $v\geq v_{\rm esc}=\sqrt{\frac{2 G M_{\rm en}(r)}{r}}$ if $r\geq r_{\rm bg}$,\\
	where $M_{\rm en}(r)$ is the total mass of gas and 
	stars enclosed in a spherical volume of radius $r$ and $v_{\rm esc}$ is 
	the velocity required by a star at $r$ to escape
	from the gas medium to spatial infinity.

It is possible that an ejected star turns around before escaping to a vacuum region if the 
instantaneous ejection velocity is not larger than the escape velocity of the whole system. 
Here, it will simply orbit the CM of the gas medium with a chance of undergoing additional 
scatterings with other stars. In this case, condition 5) would not be satisfied regardless of 
all the other conditions. In order to avoid excessively long integration times, we stop the 
simulations when $t>10^{7} - 10^{8}\yr$ (i.e., the typical life time of a star cluster, 
\citealt{Subramaniam+1995}) or the distance from the CM of any ejected star exceeds 
$10^{6}-10^{7}\AU$.

This additional distance condition can be justified as follows.   (\textit{a}) In our simulations, the 
gravitational force (or potential) from the gas dominates at $r>10^{5}\AU$ when 
$n=10^{2}\cm^{-3}$, compared to the gravitational force from the other stars in the system. 
(\textit{b}) Given that the tidal radius of a typical star cluster in the Milky Way is $< 35\pc$ \citep{Heggie2003}, 
the ejected stars will eventually escape from their host cluster. (\textit{c}) The stellar density 
in the central region of a typical star cluster is $n=10^{2}-10^{3}\pc^{-3}$, so that it 
is very likely that in such systems the ejected stars will encounter other stars in the 
cluster before moving a distance of $10^{6}-10^{7}\AU$ (the typical distance between 
stars is $\sim10^{4}-10^{5}\AU$). (\textit{d}) The peak separation between clusters in our 
Galaxy is $10^{5}-10^{6}\AU$ \citep{Subramaniam+1995}.

\section{Results}
\label{sec:results}
In this section, we present the results of our scattering experiments between two binaries
 in a background potential including the outcome probabilities and the statistical properties of 
 the final binary and single stars.

\subsection{Characterizing the effect of the background potential in different regions}
\label{result_regions}
We expect the dynamics of the stars to depend on what the dominant potential is in the region of interaction.
In general, as stars approach the system CM, the gravitational forces imparted 
by the gas become very small ($a_{\rm bg}\sim r$), since the enclosed mass 
drops to zero. In other words, in the inner region characterized by a distance 
at which the enclosed mass of gas ($M_{\rm en,gas}$) is the same as the stellar 
mass ($M_{\star}$), the stellar dynamics are primarily determined by their
mutual gravitational forces. Here, we denote this characteristic distance 
by $r_{\star}$, estimated as:
\begin{align}
\label{eq:r_star}
\frac{r_{\star}}{\AU}=\Big(\frac{M_{\star}}{4\pi\rho/3}\Big)^{1/3}\simeq 1.4\times10^{5}\Big(\frac{M_{\star}}{3\Msol}\Big)^{1/3}\Big(\frac{n}{10^{2}\cmcube}\Big)^{-1/3}\,,
\end{align}
and refer to the inner region bounded by $r_{\star}$ as \textit{region 1}.

On the other hand, stars outside \textit{region~1} are more heavily influenced by the 
background potential. Provided the same functional form for the 
potential (see equation \ref{eq4}) is adopted as for classic simple harmonic motion, 
stars can become trapped in the background potential, oscillating around the CM 
with a period of $\sqrt{\frac{3}{4\pi G \rho}}$ and an amplitude of $\sqrt{\frac{3}{4\pi G \rho}}v$
(most likely the instantaneous ejection velocity for $v$). 
The amplitude corresponds to the maximum distance reached by the stars, at which point 
the stars turn around and return toward the system CM.  We denote this return distance by 
$r_{\rm return}=\sqrt{\frac{3}{4\pi G \rho}}v$, 
and refer to the region enclosed by $r_{\rm return}$ outside \textit{region 1} as \textit{region~2}. 
The remaining volume ($r_{\rm return}<r\leq r_{\rm bg}$) is denoted \textit{region~3}\footnote{Note that in order to more precisely estimate the maximum attainable distance by the stars, we may need to add to $\sqrt{\frac{3}{4\pi G \rho}}v$ a radial distance from the system CM when the stars are completely detached from the stellar subsystem (i.e., positive relative energy between an ejected star and the remaining substellar system).}.
 But in our simulations that the distance is relatively small compared to $\sqrt{\frac{3}{4\pi G \rho}}v$.
 Figure \ref{fig:configuration} schematically illustrates the three regions. 
 The blue solid, black dotted and 
 black dashed lines correspond to \textit{region~1}, \textit{region~2} and \textit{region~3}, respectively.

In our simulations, the typical final separations between stars in a given simulation are determined by the
 termination criteria (see equation \ref{eq:tidaltoleranceparameter}), which we call $r_{\rm term}$.
This distance $r_{\rm term}$ roughly corresponds to the maximum attainable distance by the ejected stars in 
our simulations. 
To see this,
 note that upon inserting equations \ref{eq:rel_force} and \ref{eq:tid_force} into 
 equation \ref{eq:tidaltoleranceparameter}, we find: 
\begin{align}
\delta&=\frac{F_{\rm tid}}{F_{\rm rel}}\simeq\frac{2 G (m_{i1}+m_{i2})m_{j}}{r_{ij}^{3}}a(1+e) \frac{[a(1+e)]^{2}}{Gm_{i1}m_{i2}}\nonumber\\
&\simeq (2-40)\Big(\frac{a}{r}\Big)^{3}\,,
\end{align}
where the factor in front accounts for the possible range in eccentricity $e$ and mass ratio $q$ for any 
combination of bound/unbound systems ($1\leq(1+e)^{3}\leq 8$ and the term with mass 
$\sim [2-4.5]$) available to the $N=4$ case. Finally, using the adopted value for $\delta$, 
$r_{\rm term}$ is expressed as follows:
\begin{align}
r_{\rm term}&\simeq a\Big[\frac{\delta}{(2-40)}\Big]^{-1/3}\simeq 2000~a\,. 
\end{align}

If $r_{\rm term}\leq r_{\star}$, the outcomes are determined by stellar 
interactions taking place only in \textit{region~1}. 
The small tidal tolerance parameter
we take ensures that no further scatterings will occur even though the stars are most 
likely confined to \textit{region~1} until the end of the simulations.
If $r_{\star}<r_{\rm term}\leq r_{\rm return}$, 
the termination criteria are satisfied when the ejected stars reach \textit{region~2}. 
Stars in \textit{region~2} start to slow down and become trapped in the background 
potential. It is still possible that the 
stars return toward the system CM and undergo subsequent encounters prior to 
their final ejection. If $r_{\rm return}<r_{\rm term}\leq r_{\rm bg}$, the ejected stars are
completely bound to the background potential, orbiting around the system CM, causing 
subsequent scatterings with existing binaries or triples left relatively close to the system 
CM.  Assuming the semimajor axis of the final binary is comparable to the initial semimajor 
axis and $v\sim (1-2) v_{\rm cri}$ (or, roughly the escape velocity from two or three identical 
point masses), 
which is indeed what we see in Figures \ref{fig:211_x} and \ref{fig:finalv}, 
we can find expressions for $r_{\star}$, 
$r_{\rm return}$ and $r_{\rm bg}$ in terms of $a_{0}$ and $n$.  That is:

\begin{align}
	\begin{cases}
		\Big(\frac{a_{0}}{1\AU}\Big)\Big(\frac{n}{10^{2}\cmcube}\Big)^{1/3}<10^{2} \hfill\text{\textit{region~1}\,;}\\
		10^{2}<\Big(\frac{a_{0}}{1\AU}\Big)\Big(\frac{n}{10^{2}\cmcube}\Big)^{1/3}<10^{3} \hfill\text{\textit{region~2}\,;}\\
		10^{3}<\Big(\frac{a_{0}}{1\AU}\Big)\Big(\frac{n}{10^{2}\cmcube}\Big)^{1/3}<10^{5}\Big(\frac{M_{\rm gas}}{10^{10}\Msol}\Big)^{1/3}\hfill \text{\textit{region~3}\,.}
	\end{cases}
	\label{eq:threeregions}
\end{align}

Figure \ref{fig:a_n_relation} shows the three regions in the $n - a_{0}$ plane based on the 
relations derived above. We also mark all default sets of simulations (stars) 
as well as two additional sets (solid circles), but with different colors 
for each region (black for \textit{region~1}, green for \textit{region~2} and magenta for 
\textit{region~3}). 
The typical returning times (the oscillation period in $\yr$,  $\sqrt{\frac{3}{4\pi G \rho}}$) are indicated 
along the upper x-axis.

\subsection{Outcome probabilities}
\label{lown_fraction}
We classify the final end products of binary-binary scatterings according to the four outcomes:
\begin{enumerate}
	\item \,\,\ a binary and two single stars (2+1+1)
	
	\item \,\ two binaries (2+2)
	
	\item a triple and a single star (3+1)
	
	\item four single stars (1+1+1+1)
\end{enumerate}
Henceforth, we will refer to each outcome by what is given in the accompanying parentheses.

Figure \ref{fig:endproduct} shows each outcome probability as a function of 
gas density for each semimajor axis. The error bars indicate
	the Poisson uncertainties for each simulation set. The line types differentiate between the different 
semimajor axes: the solid, dot-dashed and dotted lines correspond to, respectively, 
$a_{0}=1\AU, 10\AU$ and $100\AU$. The different point types indicate the different 
outcomes: the triangles, circles, down-pointing triangles and squares correspond to, 
respectively, 2+1+1, 2+2, 3+1 and 1+1+1+1. For clarity, we use the same color 
scheme as in Figure \ref{fig:a_n_relation} for the simulations in each region.
\subsubsection{Exploring the $E_{0}=0$ neighborhood }

\begin{figure}
	\centering
	\includegraphics[width=8.9cm]{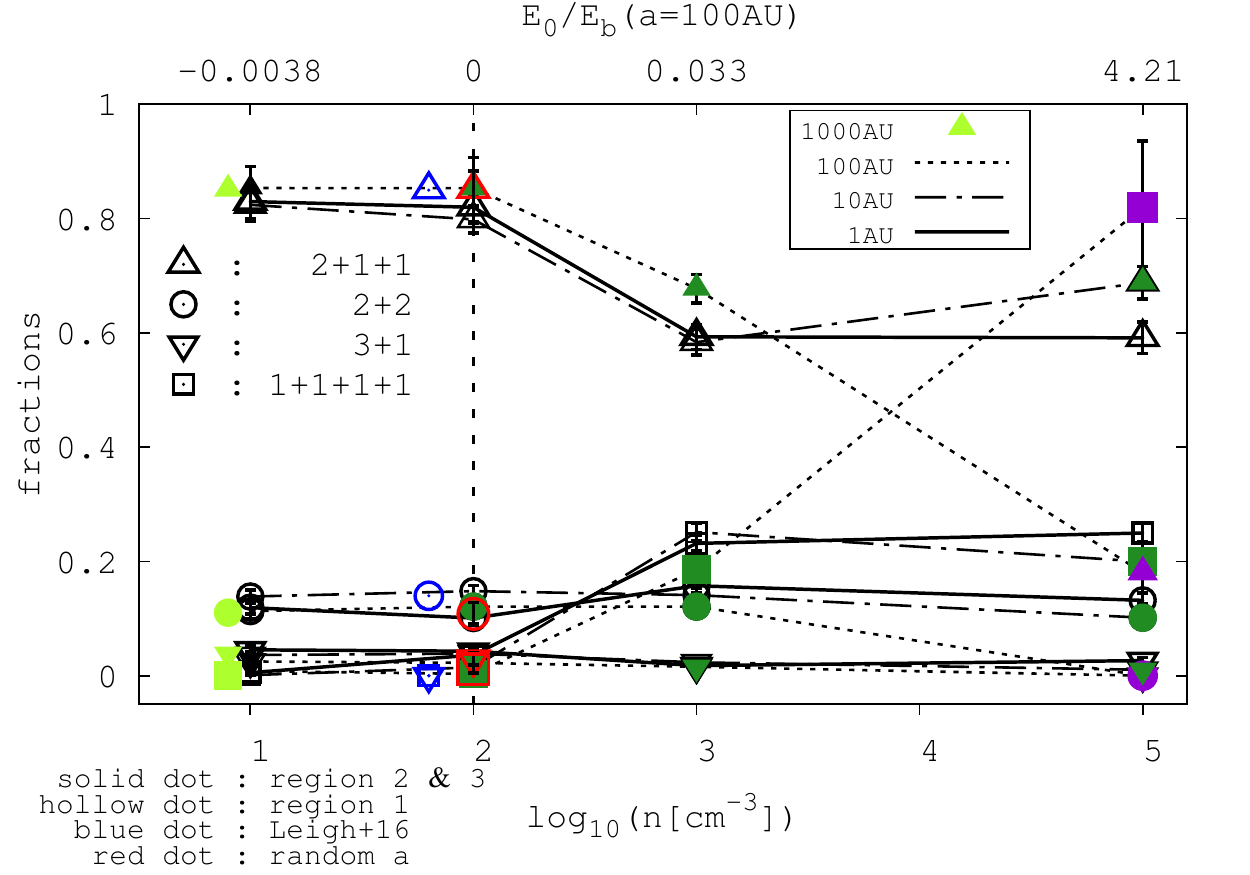}
	\caption{The fraction of each outcome as a function of gas density 
		for three discrete values of the semimajor axs. The different line types correspond
		to the different semimajor axes: the solid line, dot-dashed line and dotted line 
		represent, respectively, $a_{0}=1\AU, 10\AU$
		and $100\AU$. The different point types indicate the different encounter 
		outcomes: the triangles, circles, down-pointing triangles and 
		square dots correspond to, respetively, the outcomes 2+1+1, 2+2, 3+1 and 1+1+1+1.
		The outcome fractions at $n=0\cmcube$ and $E_{0}=0$ in 
		\citet{Leigh+16} (see Figure 4 in their paper assuming a virial ratio $k\simeq1$) are marked
		as blue dots, and those for our simulations where the semimajor axis is randomly 
		generated in the range of $a_{0} = 1 - 100\AU$ are marked as red dots. 
		The bright green dots near $n=10\cmcube$ represent the fractions 
		for an additional experiment with $n=10\cmcube$ and $a_{0}=1000\AU$.
		The  setup details are described in the text.}
	\label{fig:endproduct}
\end{figure}

Figure \ref{fig:endproduct} shows that the outcome fractions of our numerical study at 
$n=10-10^{2}\cmcube$ are almost the same as for the equivalent scattering experiments  without
a background potential (i.e., with $n=0$).
As $n$ increases from $n=10^{2}\cmcube$ to $n=10^{3}\cmcube$, the total energy
 becomes positive and the 1+1+1+1 outcome begins to appear for all semimajor axes.  
 To ensure that we have correctly calculated the outcome fractions at this critical energy limit,  we 
 run two more sets of simulations: 1) Instead of assuming 
 discrete values for the semimajor axes, 
 we generate two binaries with the same semimajor axis, but randomly generated within the
 range $a_{0}=1\AU-100\AU$. The results of this experiment (marked with red hollow dots at
 $n=10^{2}\cmcube$ in the figure) are consistent with those found adopting discrete 
 values for the semimajor axis;  
 2) we performed experiments with a density between the two critical densities, or $n\simeq300\cmcube$ 
 ($\log[n/\cmcube]=2.5$) for both $a_{0}=1\AU$ and $10\AU$. 
 As expected, the outcome fractions for $n\simeq300\cmcube$ fall between those for 
 $n=10^{2}\cmcube$ and $n=10^{3}\cmcube$ in all cases. 
 We do not show them to avoid overcrowding the figure.
 Therefore, both experiments show that the outcome fractions and their dependences 
on the gas density are consistent with the ranges and values physically allowed 
 by the total energy (e.g., the 1+1+1+1 
 outcome only occurs for positive total energies). 

We compare our results with the
outcome probabilities at $n=0\cmcube$ and $E_{0}=0$ given in \citet{Leigh+16}
(see Figure 4 at virial ratio $k\simeq1$). These are marked as blue hollow dots using the 
same shape for each outcome. To avoid overlapping with other data points, 
these points are marked off the $E_{0}=0$ line 
(the vertical dotted line at $n=10^{2}\cmcube$). 
Note that the numerical scattering experiments in \citet{Leigh+16} agree with those 
shown here for $a_{0}=1\AU$ since all assumptions are the same, 
including the semimajor axis ($a_{0}=1\AU$), 
the eccentricity ($e=0$), the stellar mass ($M=1\Msol$) and the total initial energy ($E=0$). 
Additionally, we find that the outcome fractions in our study at $n=10-10^{2}\cmcube$ are 
still comparable to those found in \citet{Leigh+16}, even for our other choices of the 
semimajor axis ($a_{0}=10\AU$ and $100\AU$). 
These results are consistent with our expectations 
in \textit{region~1}. In this region, the gravitational forces 
between stars dominate, and hence the 
stellar dynamics should remain the same as in the absence of the background potential.

\begin{figure}
	\centering
	\includegraphics[width=8.9cm]{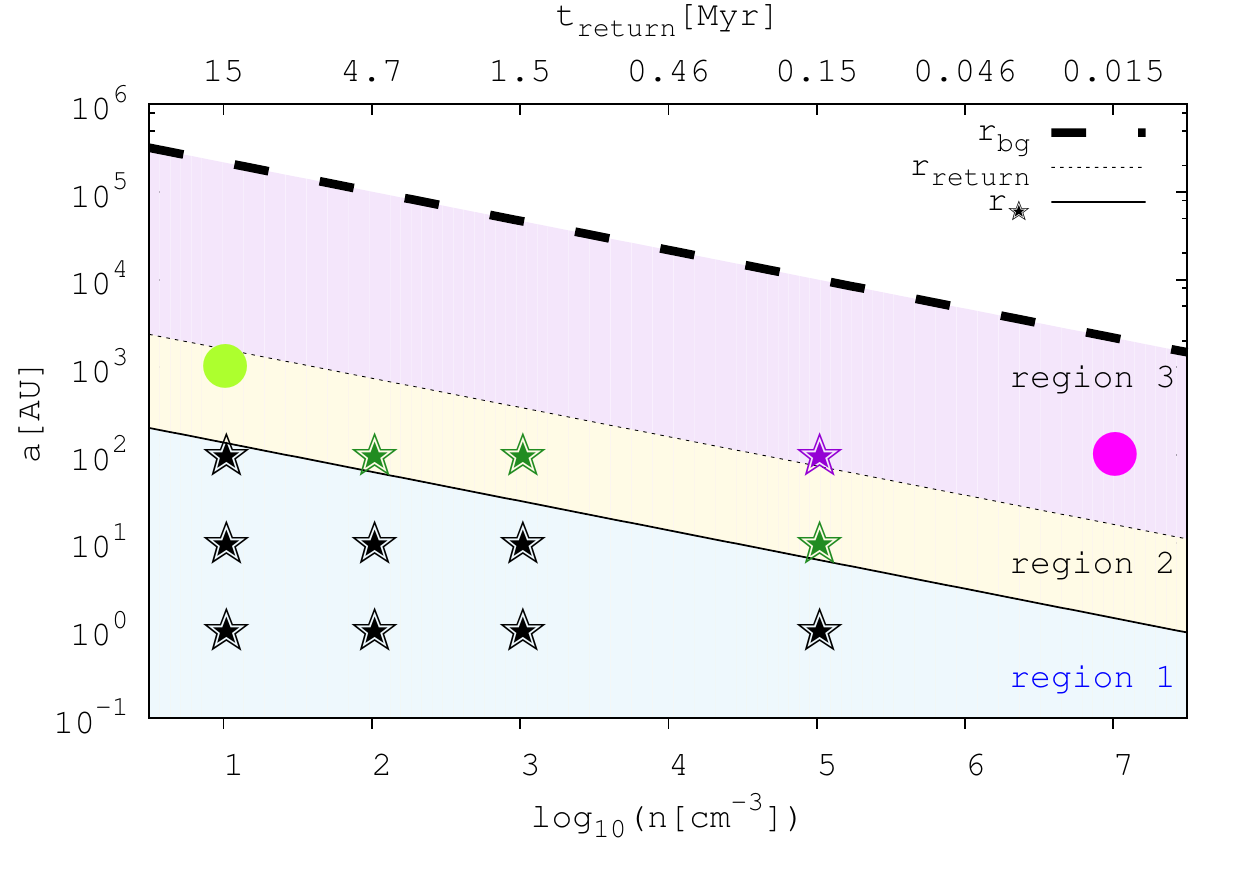}
	\caption{A schematic diagram projected on to the 2-D $n - a_{0}$ plane showing three distinct 
		regions. The relations between $n$ and $a_{0}$ defining the outer boundary of each 
		region are given in Equation \ref{eq:threeregions}. We mark all default sets of 
		simulations (stars) as well as two additional sets (solid circles), but with different colors 
		for each region (black for \textit{region1}, green for \textit{region2} and magenta for 
		\textit{region3}). The typical returning times (the oscillation period in$\yr$,  
		$\sqrt{\frac{3}{4\pi G \rho}}$) are indicated along the upper x-axis.  }
	\label{fig:a_n_relation}
\end{figure}

However, as the density increases,  the final encounter outcomes 
and their probabilities change, due to the effect of the background potential,
which acts to reset the zero-point of the total system energy.
The fractions, especially for $a_{0}=100\AU$, diverge slightly from 
	the results for the other semimajor axes as $n$ increases.  We start to see that the ejected stars 
	return toward the system CM in 
	\textit{region~2} (see Figure \ref{fig:a_n_relation}).  Consequently, the stars take part in repeated 
	scattering events that would  not
	have occurred in the absence of a background potential.  This leads 
	to discrepancies in the final outcome probabilities relative to the no background potential case. 
	This regime is different from the 
	case where stellar interactions dominate, as it happens in \textit{region~1}.  Similarly, we 
	find (see Figure \ref{fig:finalv}) that the final velocity distributions for the 
	ejected stars are shifted to lower velocities for $a_{0}=100\AU$.  This could be 
	because: 1) it is more likely that the simulations stall when the ejected stars slow down provided 
	$r_{\star}<r_{\rm term}<r_{\rm return}$ in \textit{region~2};  2) in some cases the 
	simulations reach a pre-set maximum computation time.  In this case, slower velocities 
	(a snapshot or instantaneous velocity sampled randomly over the entire oscillatory motion) 
	happen to often be 
	recorded as the final velocity. In order to understand if this only occurs for the 
	$a_{0}=100\AU$ case (largest initial value in the default range of densities) at the critical energy limit, 
	we run another simulation with an even larger semimajor axis, $a_{0}=10^{3}\AU$, 
	and negative total energy ($n=10\cmcube$).  In this case, we might expect a similar 
	behavior for the ejected stars in \textit{region~2}. As before, we clearly see returning stars
	 in this experiment. However, as marked by the green solid dots slightly off $n=10\cmcube$ 
	 (to avoid overlapping) in Figure \ref{fig:a_n_relation}, the cases with smaller semimajor axes 
	 have almost the same outcome probabilities. 
	 This is because the time taken for the ejected stars to come back and 
	 engage in another scattering event (i.e, the oscillation period
	 indicated along the upper x-axis in Figure \ref{fig:a_n_relation}) is quite long.  Hence, 	
	 the number of encounters is insufficient to have a noticeable effect 
	 before the maximum simulation time is reached.  Note that the total computation time includes
	the initial drop-in time at the onset 
	 of the simulations and the duration of the interaction before the final ejection event occurs.
	 This can also help to explain why the outcome probabilities 
	 diverge more for higher $n$; that is, higher encounter frequency.
 	 From these experiments, we suggest that 
	 the depth of the potential is one of the essential factors in determining deviations from isolated 
	 stellar dynamics.

\begin{figure*}
	\centering
	\includegraphics[width=8.4cm]{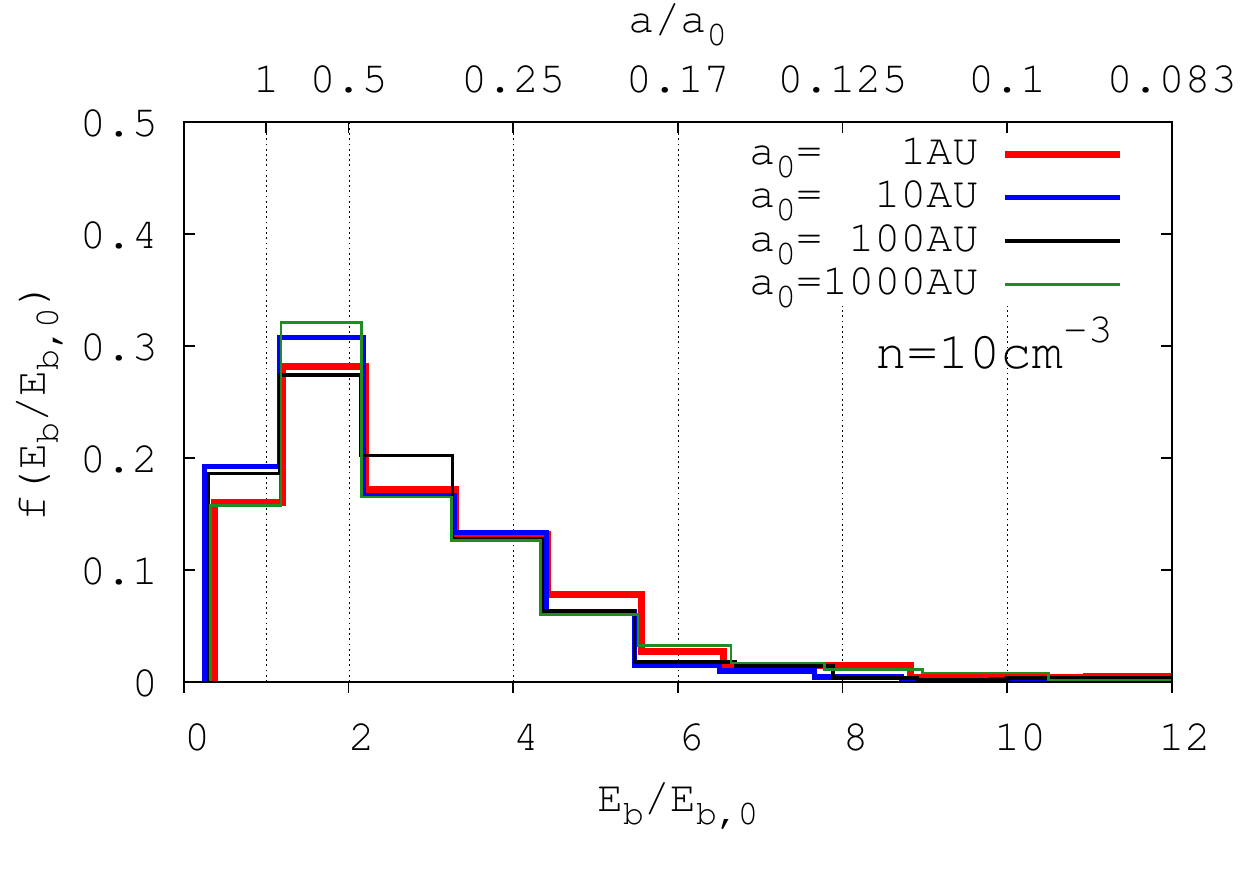}
	\includegraphics[width=8.4cm]{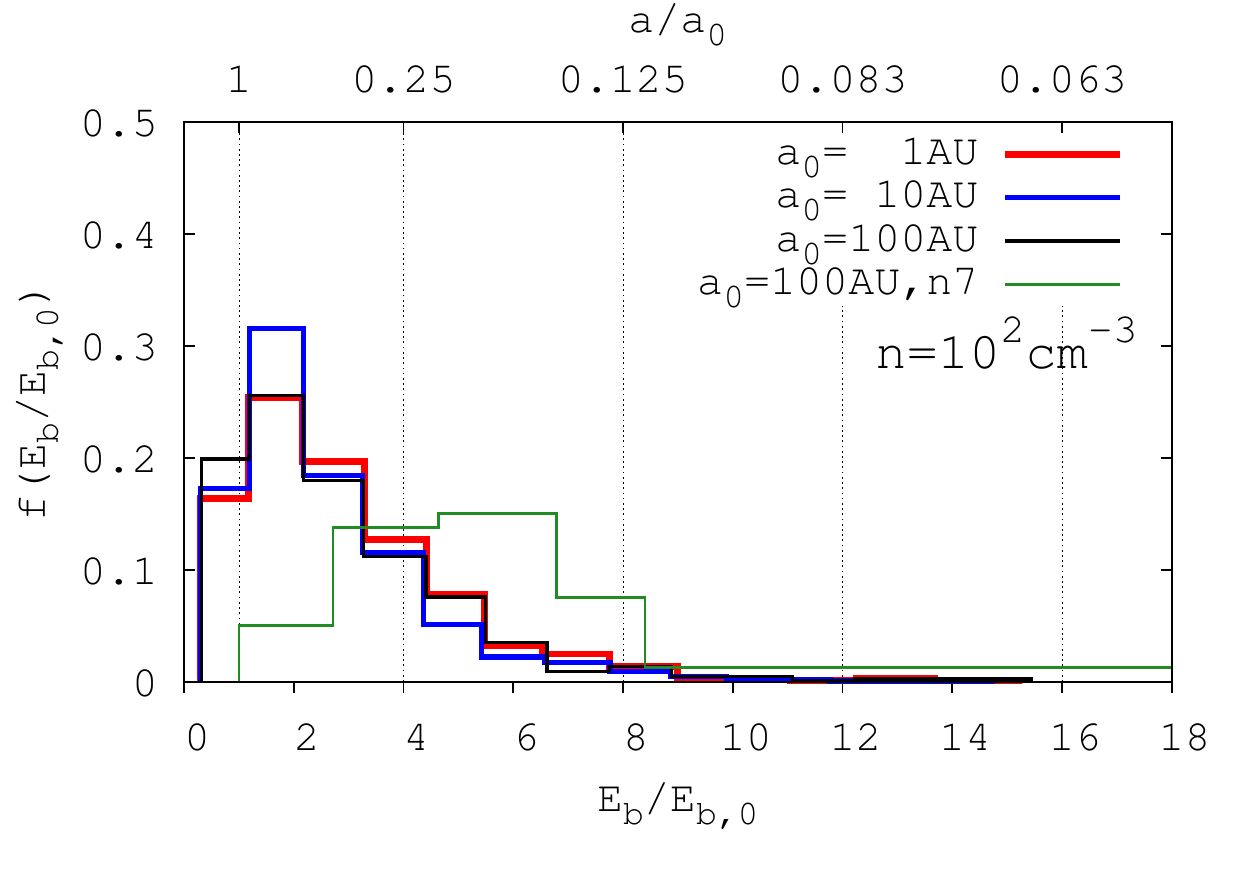}
	\includegraphics[width=8.4cm]{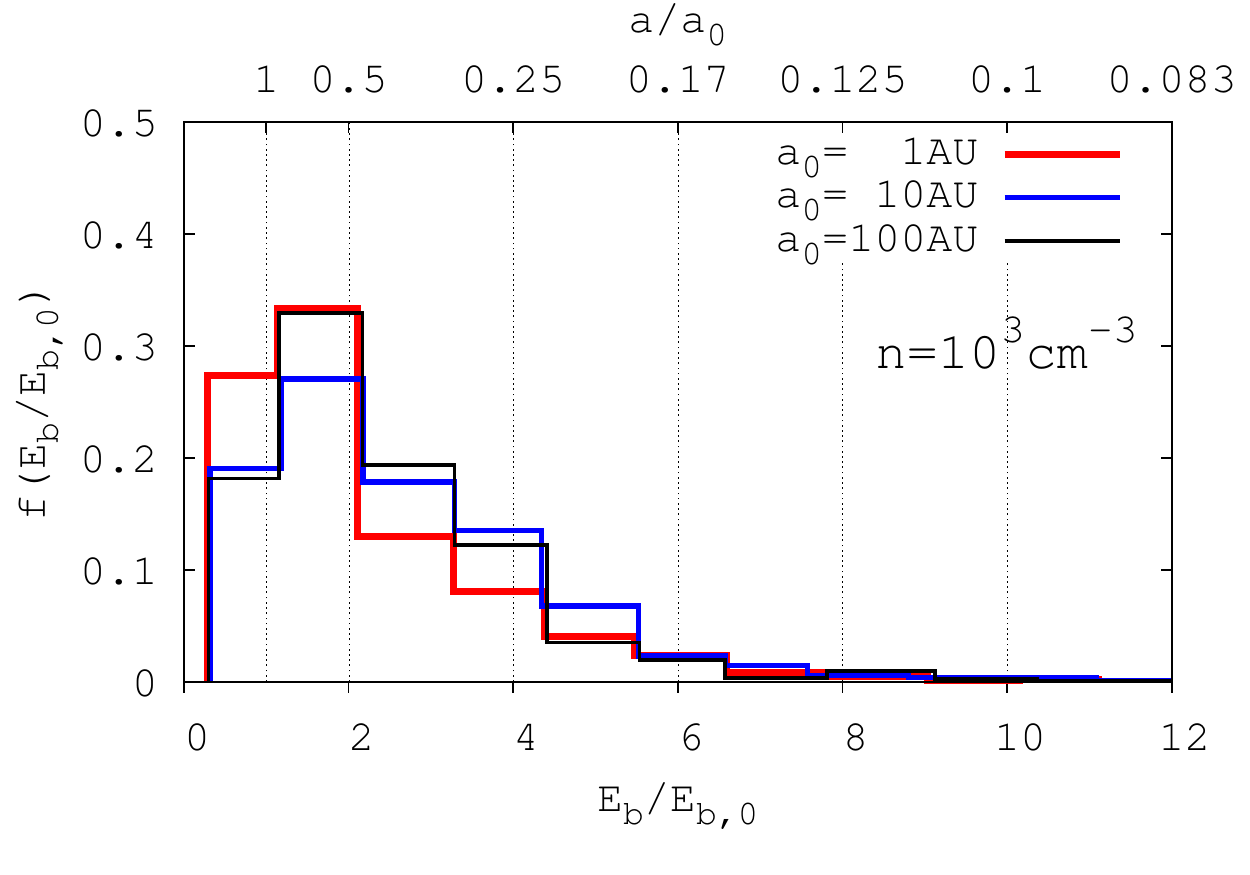}
	\includegraphics[width=8.4cm]{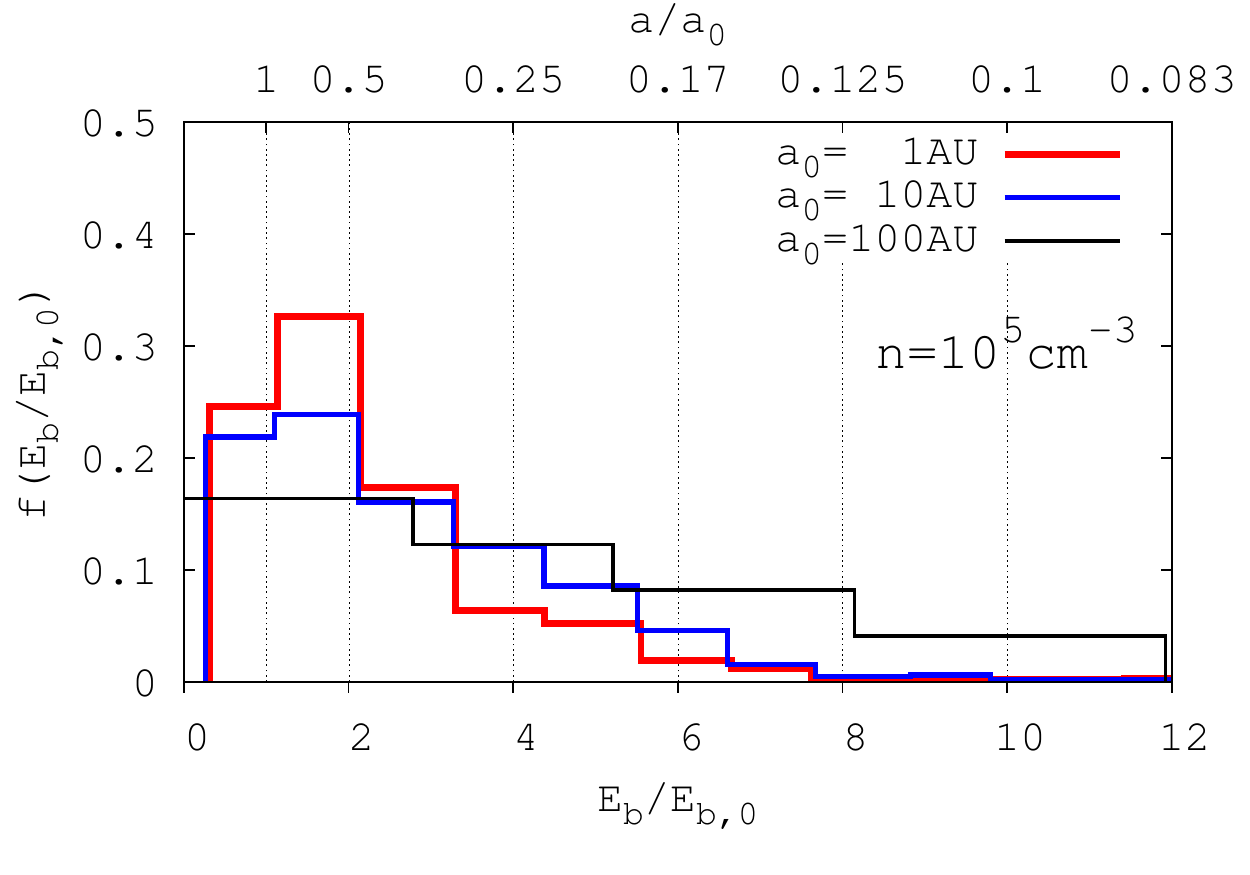}
	\caption{The binding energy distributions (scaled by the initial binary binding energy) 
		for binaries formed during 2+1+1 outcomes, 
		for number densities of $n=10\cmcube$ to $n=10^{5}\cmcube$ 
		(from the \textit{top left} to \textit{bottom right} panels, forming a ``Z" configuration). The red, blue and black solid lines correspond to, respectively,
		$a_{0}=1\AU, 10\AU$ and $100\AU$. Particularly, we also show the distribution 
		for $n=10^{7}\cmcube$ (green line) in the panel for $n=10^{2}\cmcube$ since 
		they have same total energy. The semimajor axis corresponding to a 
		given binding energy is shown along the upper axis. We provide the best fits for $n=10-10^{3}\cmcube$ in 
		Table \ref{tab:211_Ebfitting}.}
	\label{fig:211_x}
\end{figure*}

Interestingly, just above the zero-energy limit, the fraction of 2+1+1 outcomes decreases by 
about the same amount that the 1+1+1+1 outcome increases, whereas the fractions of the other two outcomes 
(3+1 and 2+2) remain the same. 
The probability of having a 1+1+1+1 outcome appears to be determined by 
whether or not all four stars are promptly ejected. In other words, 
if the stars are ejected one at a time, then a 1+1+1+1 outcome does not occur. 
This is because once a single star is ejected with positive kinetic energy, 
then the left-over system should have
a negative total energy, which can never end up fully ionized (by energy conservation). 
Note, however, that if the initial total energy is positive and sufficiently large, 
it is possible that after an ejection event the remaining subsystem still 
has a positive total energy, and so can be reionized into single stars later on. 
If $n=10^{3}\cmcube$, then the total energy is much smaller than the typical 
kinetic energies of ejected single stars (estimated as $v_{\rm esc}\sim v_{\rm cri}$).  Thus, 
we expect that the prompt ejection of all four stars is the only channel for full ionization.
 As the 1+1+1+1 outcome begins to appear, the fractions of all other outcomes 
 decrease (by what appears to be a roughly constant fraction). Since the 
 2+1+1 outcome probability is the highest, its decrease 
 looks especially significant. 

One other point to note in the density range $n=10\cmcube$ to $n=10^{3}\cmcube$ 
is that the increase we find in the probability of having a 1+1+1+1 outcome 
(i.e., from 0 to 0.2) is relatively large compared to the results of other scattering experiments
\citep[i.e][]{Mikkola1983,Fregeau+2004}. This discrepancy is most likely due to 
the presence of the background potential, but the different choices for 
the impact parameter and eccentricity could also contribute. In our study, we only 
consider initially circular orbits and zero impact parameters (i.e., head-on collisions).  
However, other studies draw the eccentricity and impact parameter from specified distributions.  
For instance, when two binaries collide head-on there is a higher probability of full 
ionization occurring relative to encounters with non-zero impact parameter, since all four 
stars are more likely to interact strongly at the first encounter such that 
the two binaries would both be (at least temporarily) 
fully ionized. Therefore, initial configurations such as 
head-on collisions can increase the fraction of 1+1+1+1 outcomes. 
Meanwhile, the background potential serves to enhance the effects of gravitational 
focusing, drawing all stars toward the system CM.

\subsubsection{Exploring the strong-background potential, positive-energy regime}
To explore the dynamical effects when the background potential is very strong, we
further ran simulations with $n=10^{5}\cmcube$ .
As shown in Figure \ref{fig:a_n_relation}, each set of simulations at $n=10^{5}\cmcube$ 
corresponds to a different region. From this we expect that the results from each 
set of simulations should reflect the characteristic behavior of the stars in the 
corresponding region. And this is exactly 
what we see from the outcome probabilities in Figure \ref{fig:endproduct} and 
the binary binding energy distributions in the \textit{bottom right} panel of Figure \ref{fig:211_x}.

The outcome probabilities become different for each semimajor axis. 
	This is different from the cases with $n=10- 10^{3}\cmcube$ in which each outcome probability for all the semimajor axes consistent within one or two standard deviations as long as $n$ is same.
More explicitly, for $a_{0}=1\AU$, these probabilities are 
almost the same as for $n=10^{3}\cmcube$. Since for this value of $a_{0}$ 
 the gravitational forces between the stars still dominate 
 over the background potential (\textit{region~1}), 
the outcome probabilities are expected to remain the same as those for $n=0$.
Since the cross section for each outcome does not 
vary much \citep{Sweatman2007,Heuvel1992} from 
$n=10^{3}\cmcube$ to $n=10^{5}\cmcube$ (0 - 0.04 in the unit of 
$E_{\rm b}(a=1\AU)$), these outcome probabilities are reasonable. 
Notice that the units shown along the upper x-axis of the figure correspond to $E_{\rm b}(a=100\AU)$. 
For $a_{0}=10\AU$, the outcome probabilities start to differ, similar to the cases for
$n=10^{2}$ and $10^{3}\cmcube$ cases with $a_{0}=100\AU$ (\textit{region2}). 

Interestingly, for $a_{0}=100\AU$, 
the outcome probabilities for 2+1+1 and 1+1+1+1 reverse and 
all other outcomes (3+1 and 2+2) are completely suppressed. 
For the cases with the 2+1+1 outcome, we find that subsequent stellar scatterings continue, 
but more frequently due to the stronger potential and hence higher accelerations, 
until two single stars are completely ejected from the system. 
Once a compact binary forms with sufficiently large binding
energy, 
it survives successive encounters, 
becoming more and more compact via the slingshot mechanism. This also implies that it is more 
likely that a wide binary or triple will be ionized (i.e., the complete suppression of the 2+2 and 
3+1 outcomes). In the presence of a deep potential, a single star has to gain a higher escape velocity for
 complete escape, necessarily leading to a hardened binary. Recall that 
 $v_{\rm esc}\sim\sqrt{\rho}$ (see the additional termination criteria 6). As a result, we
 see a shift of the binding energy distribution towards the higher energy range 
 in the \textit{bottom right} panel of Figure \ref{fig:211_x}. 
Otherwise, like the cases with the 1+1+1+1 outcome, two 
 binaries should be first ionized into four stars in order to escape.

We performed one additional experiment 
to more clearly understand the effects of the background potential on the 
subsequent stellar dynamics for negative total energies. 
In this experiment, we assume $a_{0} = 100\AU$ and $n=10^{7}\cmcube$. 
With these choices, we are able to find a system corresponding to \textit{region 3} 
where the total energy is the same 
as for the case with $a_{0}=100\AU, n=10^{2}\cmcube$ (i.e. $E_{0}\simeq 0$),
by adjusting $v_{\rm rel}$ and $r_{12}$. For this particular experiment, we set 
$M_{\rm gas}=40\Msol (r_{\rm bg}=6600\AU)$.
This choice for the total gas mass is such that 
stars can escape from the background potential well 
and the effects of the background potential should be significant.
In this experiment the only end product is 2+1+1 (cf. 2+1+1 and 1+1+1+1 in the 
simulation for $E_{0}>0$ with $n=10^{5}\cmcube$ and $a_{0}=100\AU$).
 It is possible that two 
binaries are fully ionized but given the total energy budget available to all 
four single stars, they cannot have sufficiently high escape velocities
(or kinetic energies) to meet the criterion for all four single 
stars to pass the outer potential boundary. 
We present the binding 
energy distribution for this case (green solid line) in the \textit{top right} panel of
Figure \ref{fig:211_x} ($n=10^{2}\cmcube$)  
since they share the same total energy $E_{0}$. 
The distribution is concentrated more toward higher binding 
energies, similar to the $n=10^{5}\cmcube$ and $a_{0}=100\AU$ case 
(black solid line in the \textit{bottom right} panel). In addition, 
the final binaries are always more compact than the initial binaries
(the maximum semimajor axis for the final binary is $a=64\AU<a_{0}$ or, $E_{b}/E_{b,0}>1$).

\subsection{Statistical properties of the scattering products}
	
In this section, we describe the statistical properties of the binaries and single stars formed 
in our simulations in the presence of a background potential for both the 2+1+1 
and 2+2 outcomes. That is, we show the final distributions of binding energies 
(semimajor axes) and eccentricities for all binaries formed in our simulations, as well 
as the final relative velocities of the ejected single stars and the binary for the 2+1+1 outcome, 
and between the two binaries in the 2+2 outcome.  We provide fitting formulae for all distributions, 
except for the velocity distributions, in Appendix \ref{appendix:fitting}.

\subsubsection{The 2+1+1 outcome}
\label{results_211outcome}

\begin{figure}
	\centering
	\includegraphics[width=1.0\linewidth]{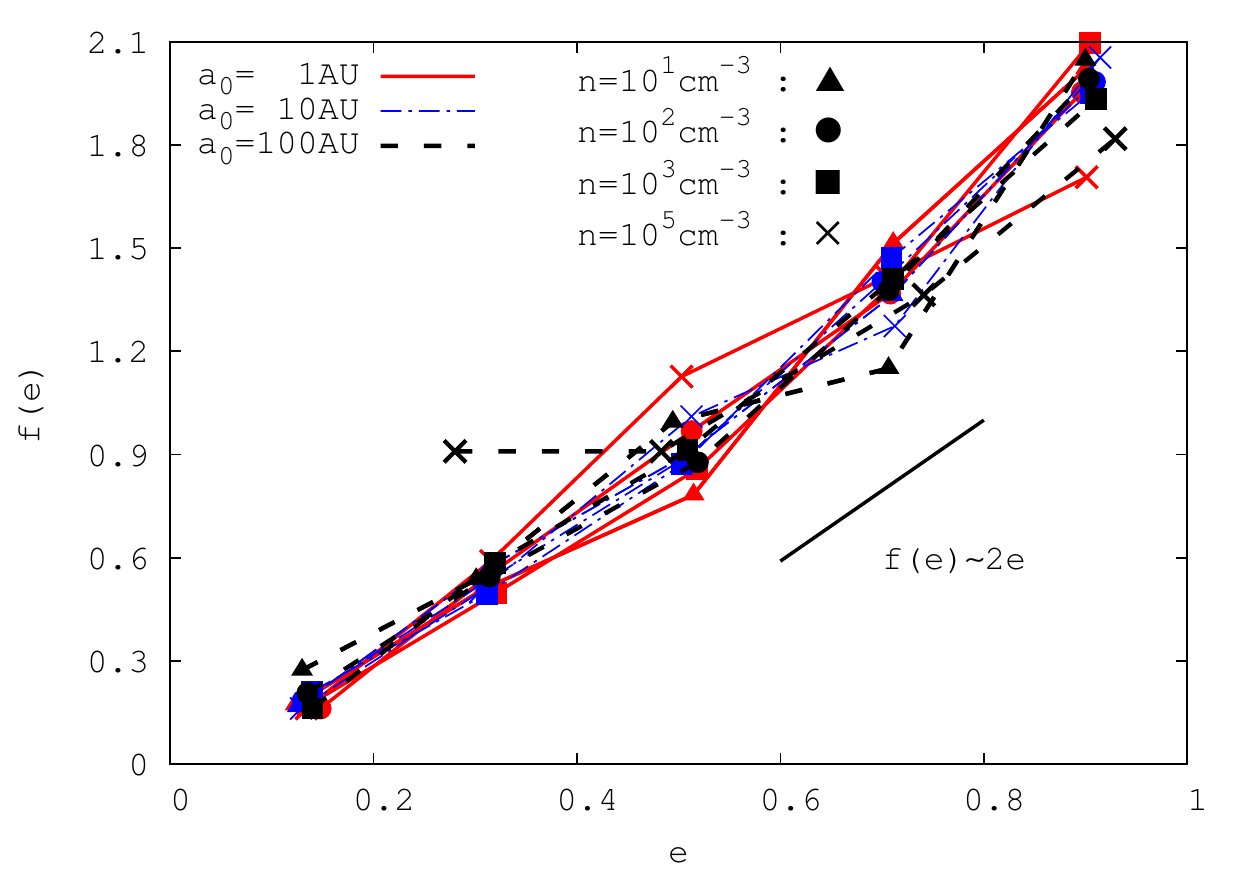}
	\caption{The eccentricity distributions for binaries formed during 2+1+1 outcomes, 
		for all number densities and semimajor axes. The same 
		colors are used as in Figure \ref{fig:211_x}. All distributions 
		follow a thermal distribution function, $f(e)\sim 2e$, 
		except for $n=10^{5}$ with $a_{0}=100\AU$, 
		the set corresponding to \textit{region~3}).}
	\label{fig:211_e}
\end{figure}

\begin{figure*}
	\centering
	\includegraphics[width=8.4cm]{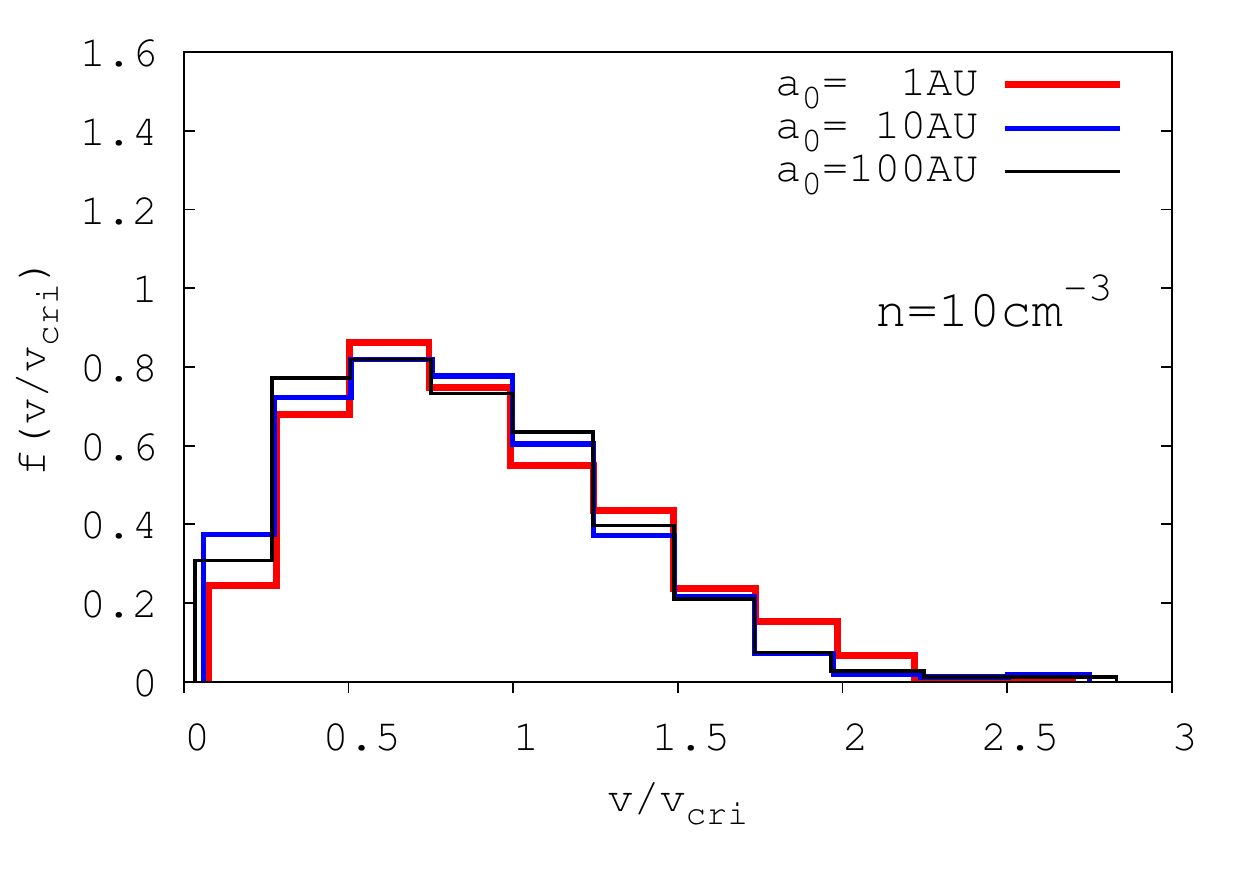}
	\includegraphics[width=8.4cm]{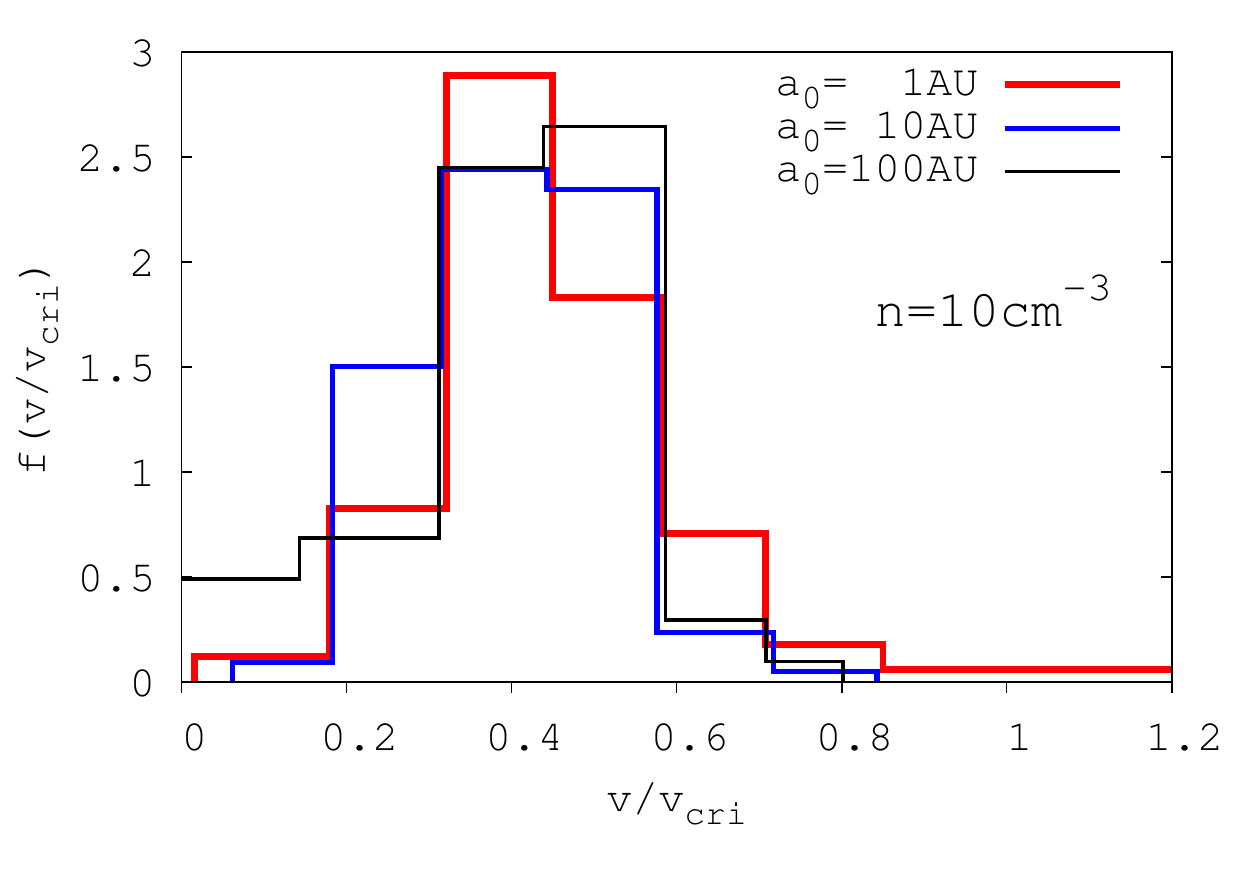}
	\includegraphics[width=8.4cm]{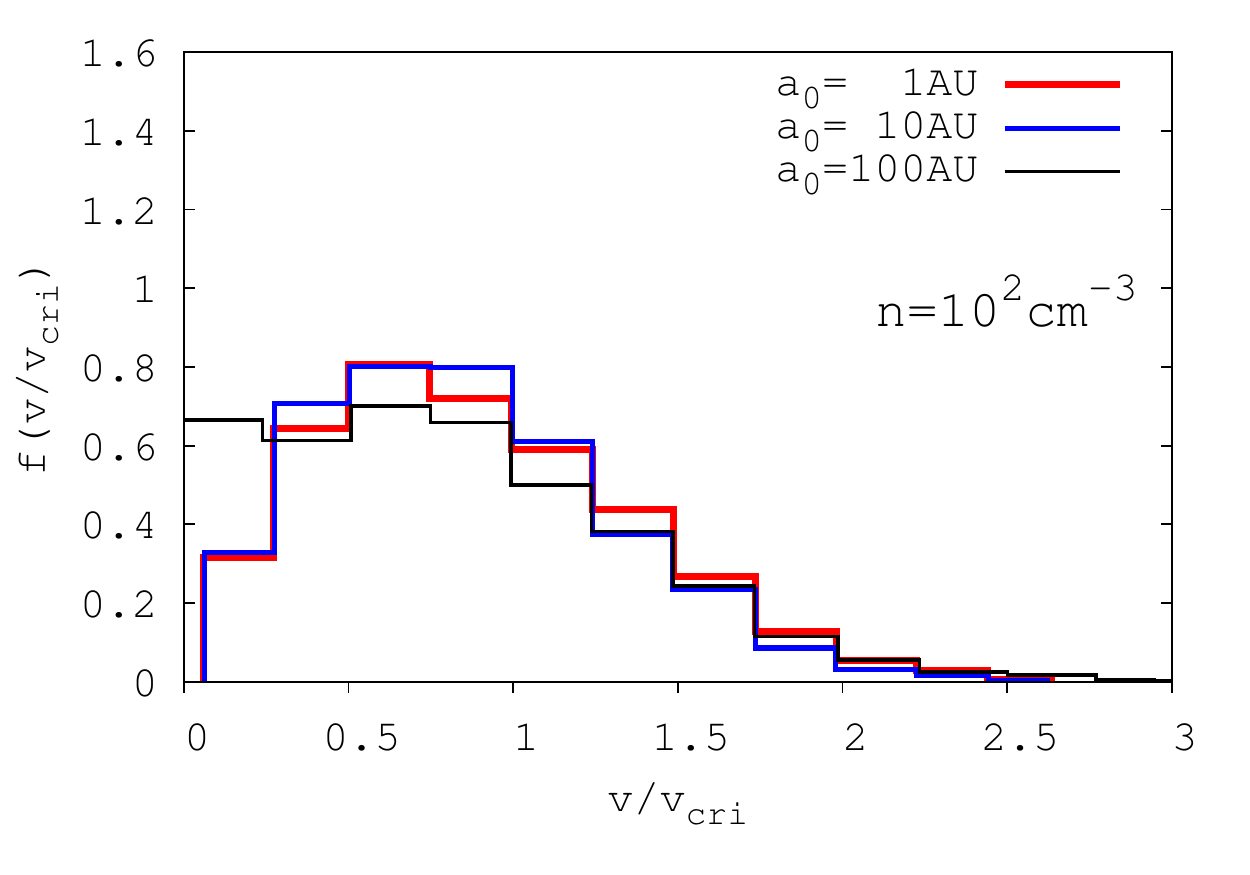}
	\includegraphics[width=8.4cm]{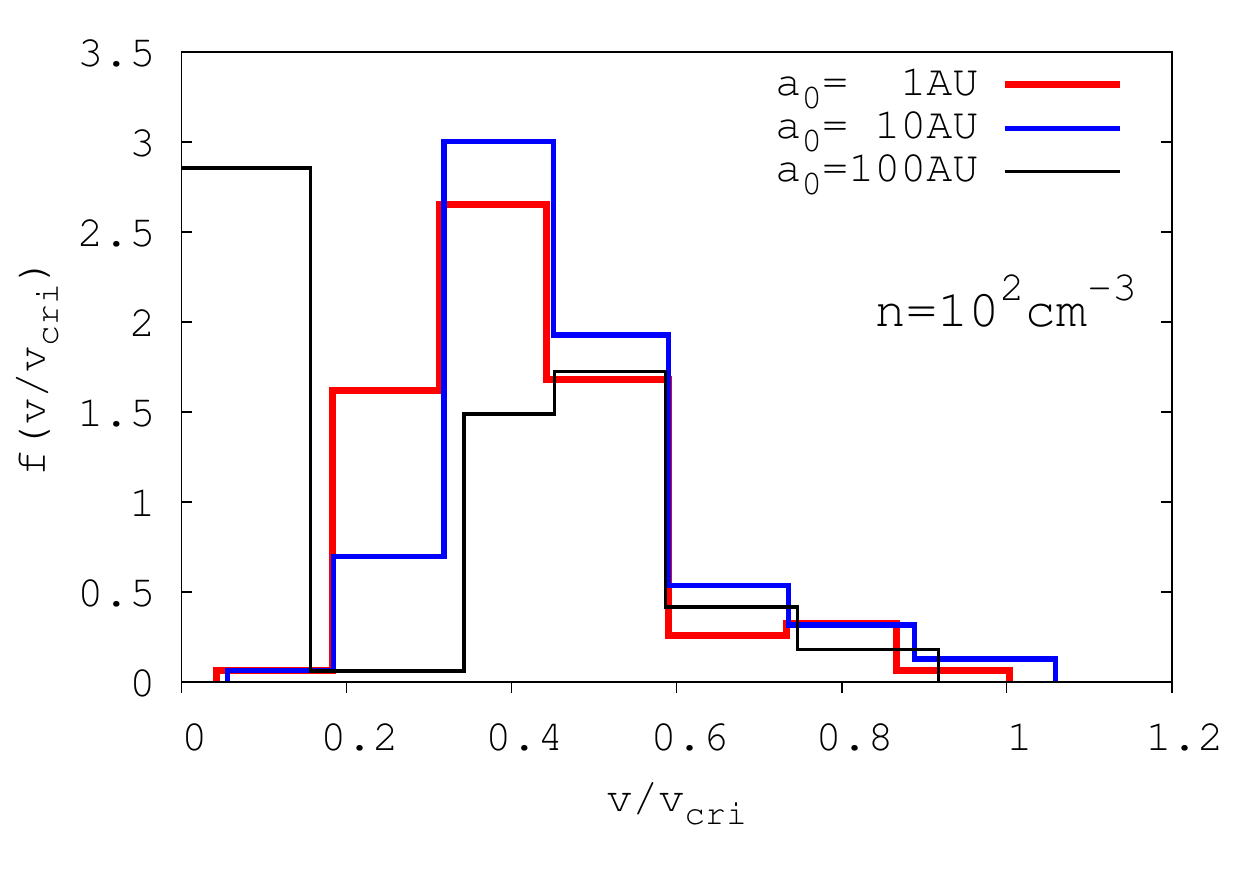}
	\includegraphics[width=8.4cm]{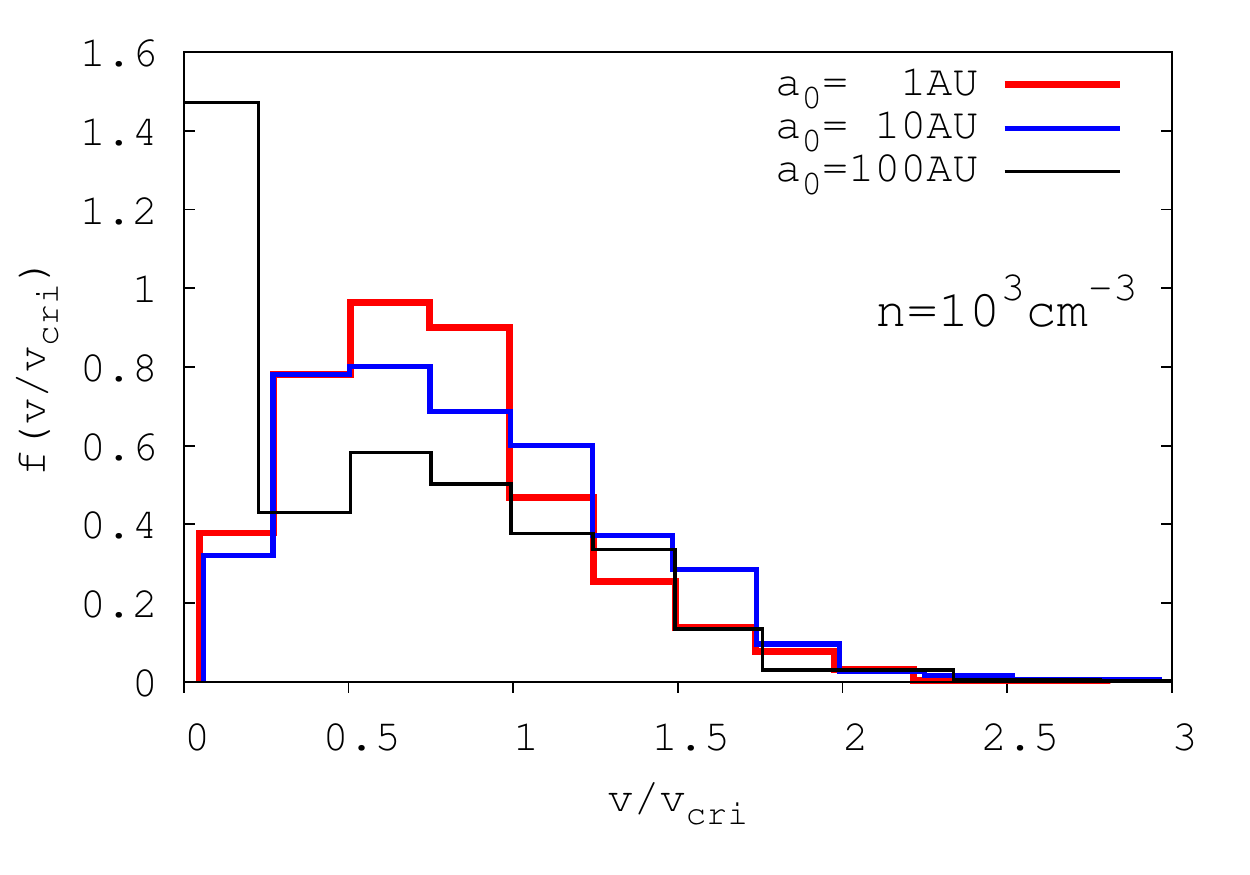}
	\includegraphics[width=8.4cm]{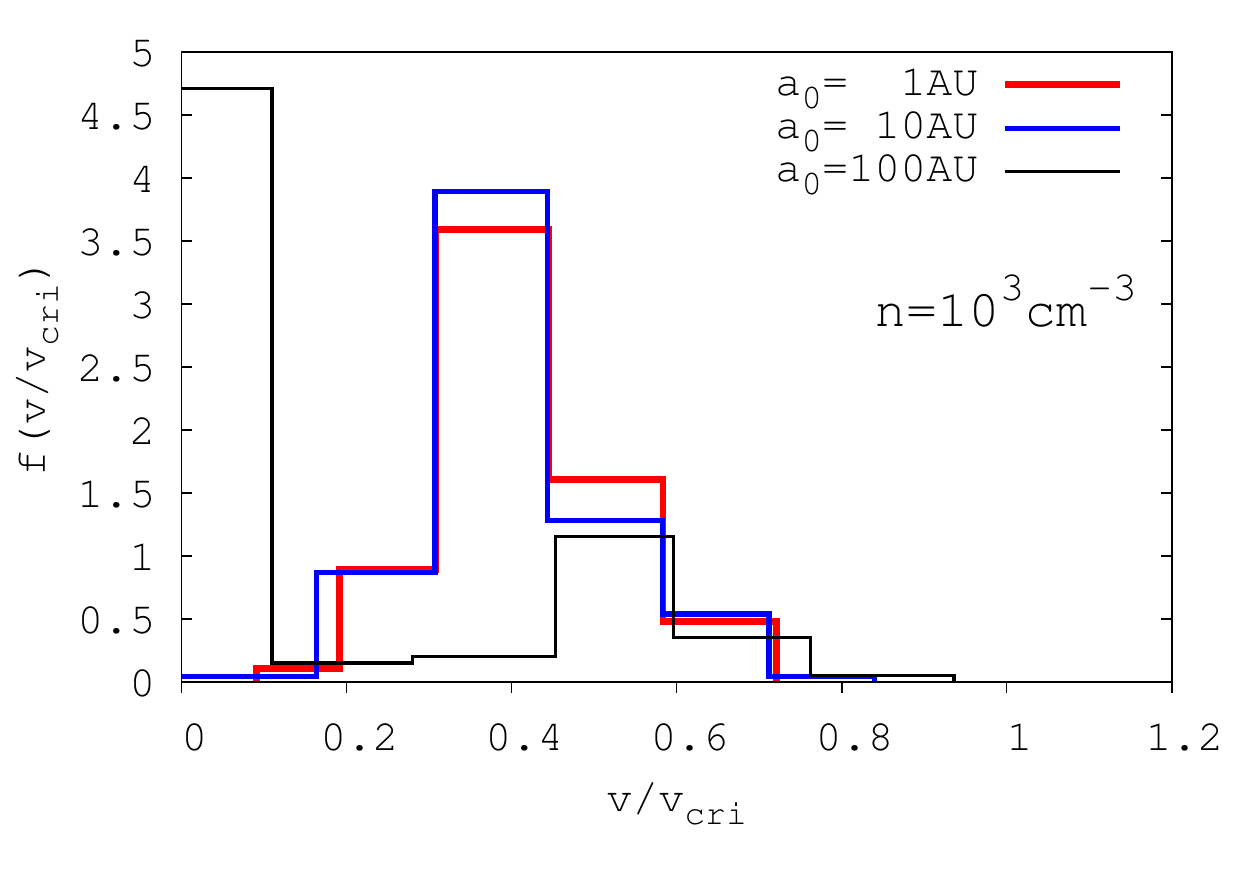}
	\caption{The final velocity distributions (scaled with $v_{\rm cri}$) 
		for binaries formed during 2+1+1 outcomes (\textit{left} column) 
		and 2+2 outcomes (\textit{right} column), 
		for number densities of $n=10\cmcube$ to $n=10^{3}\cmcube$ 
		(from \textit{top} to \textit{bottom} panel). In both panels, the same colors 
		are used as in Figure \ref{fig:211_x}.
		}
	\label{fig:finalv}
\end{figure*}

The 2+1+1 outcome has the highest probability of occurring, relative to the other outcomes.  
Unlike during full ionization (i.e., 1+1+1+1), the single star ejection events 
are not necessarily prompt. Typically, both binaries are temporarily ionized at first impact 
and one single star is ejected, leaving behind an unstable triple system, 
followed by the eventual ejection of the second single star. For the 2+1+1 outcome, the 
final total energy may be regarded as consisting of two components, namely the (negative) 
binding energy of the final binary and the (positive) kinetic energies of the two single stars 
and the binary (with respect to the system CM and ignoring the gravitational potential between 
the escaped objects). Since we are exploring the nearly zero energy limit, the final binding 
energy of the binary immediately tells us about the kinetic energies of the ejected single stars.

\begin{figure}
	\centering
	\includegraphics[width=8.3cm]{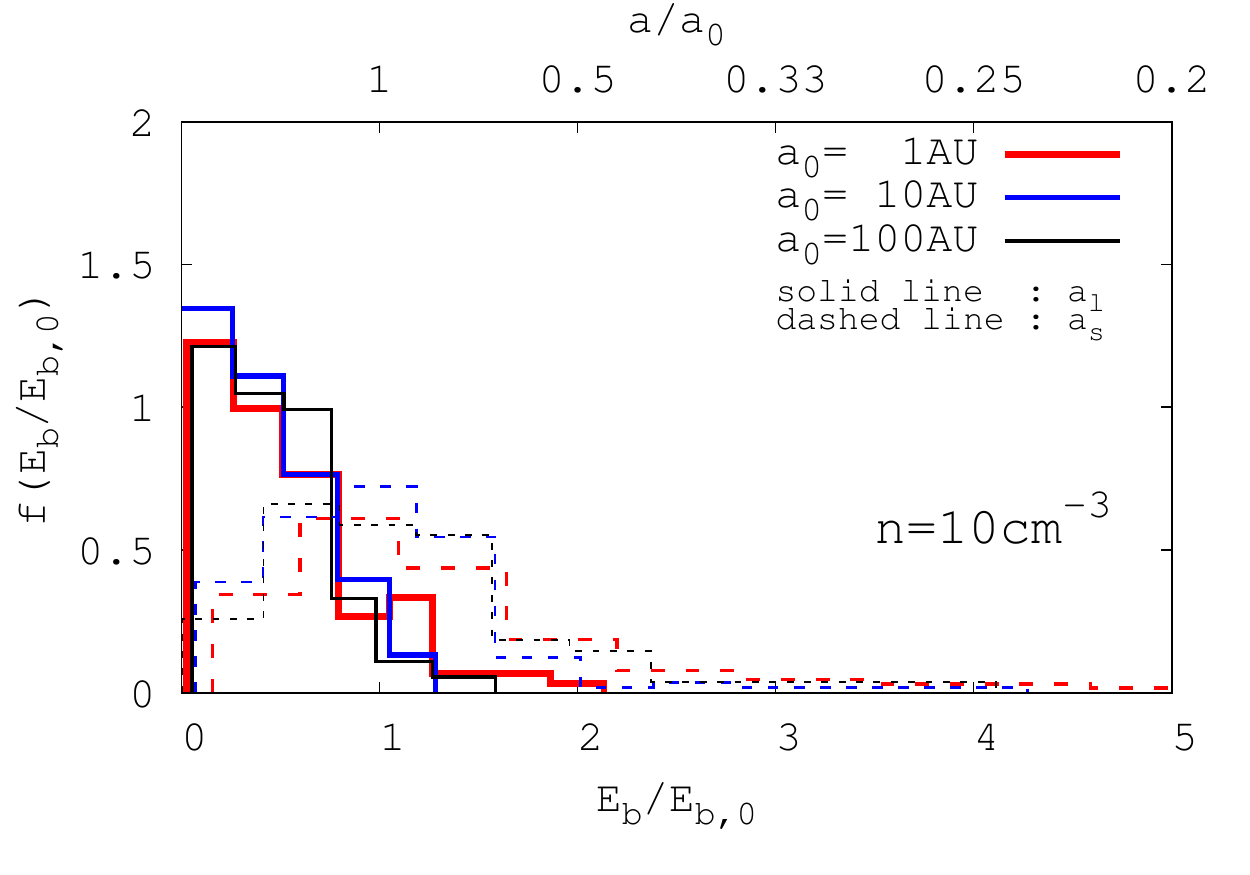}
	\includegraphics[width=8.3cm]{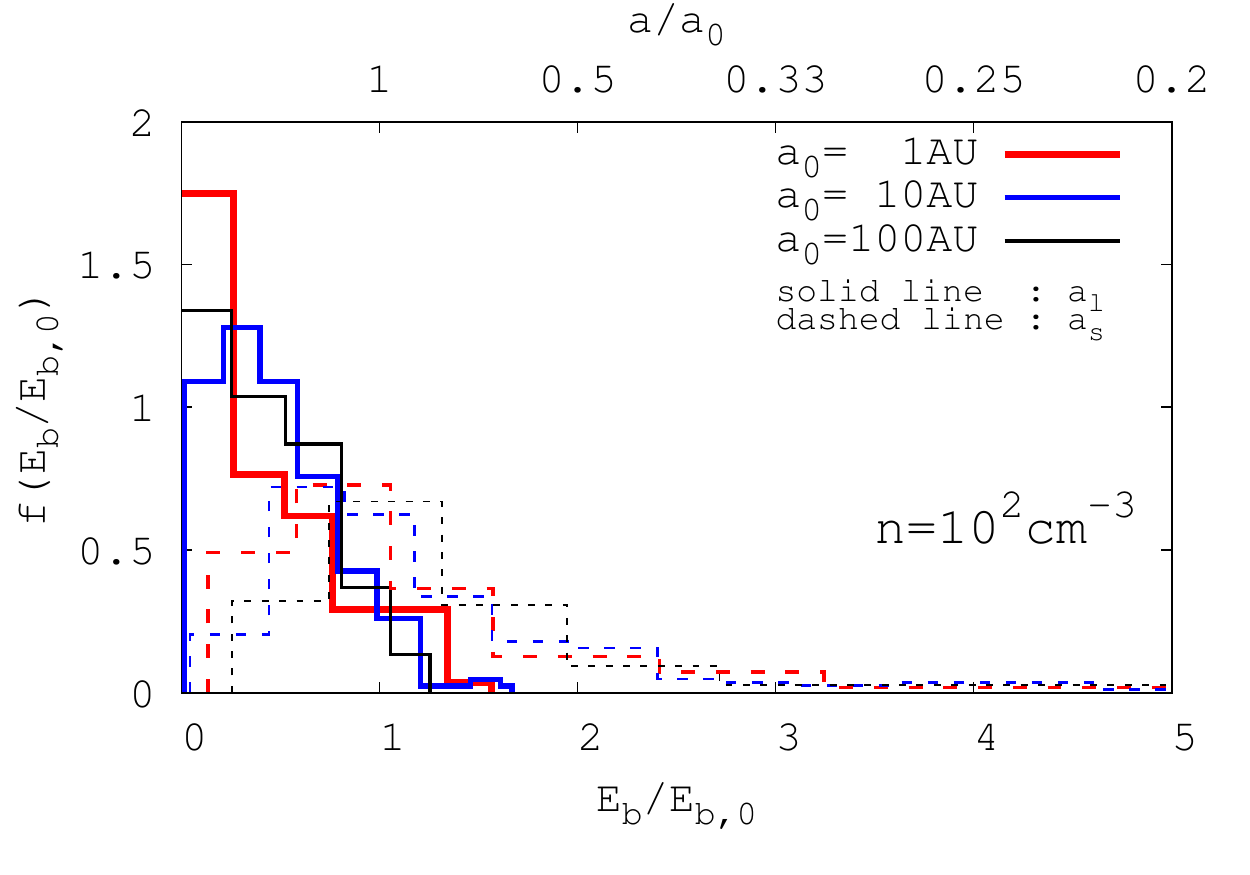}
	\includegraphics[width=8.3cm]{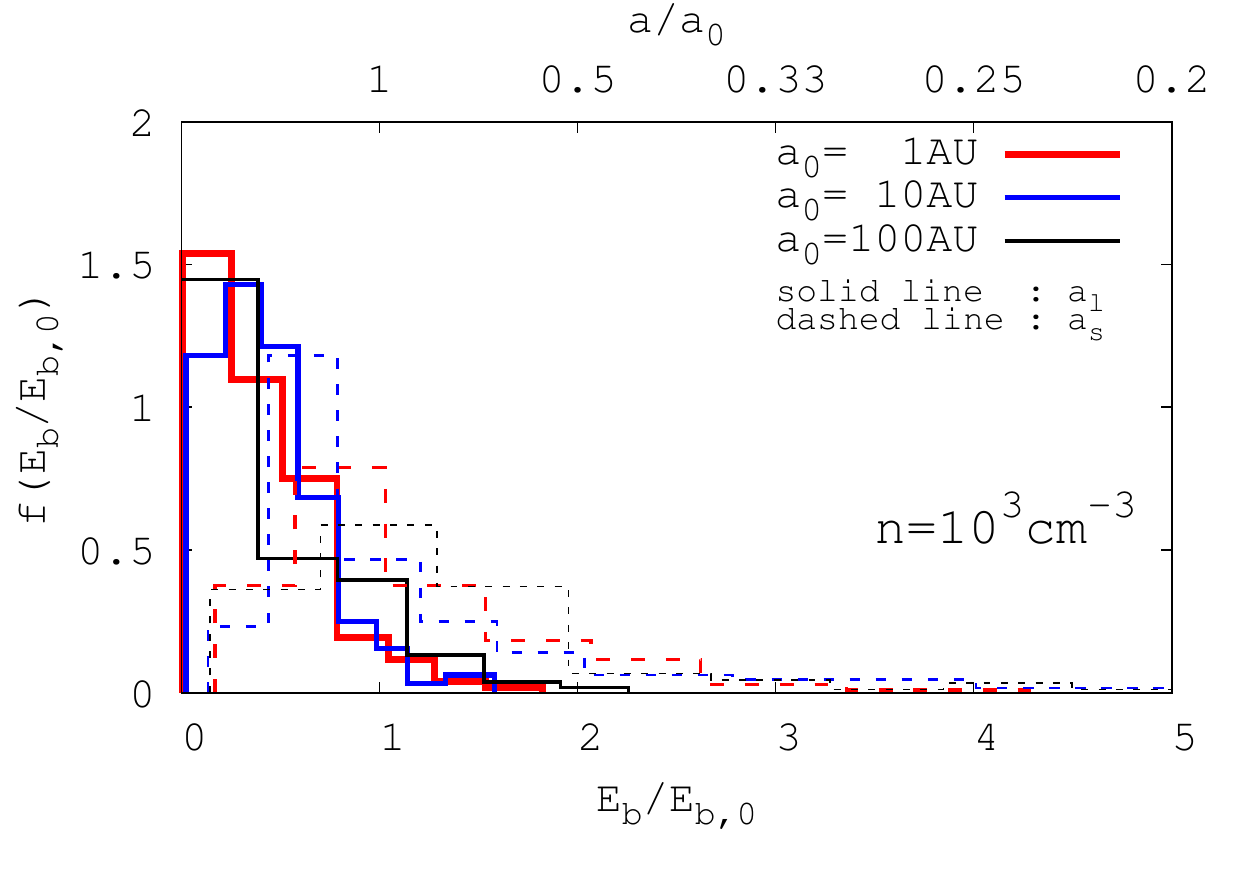}
	\caption{The final binary binding energy distributions for both binaries produced 
		during 2+2 outcomes for number densities of $n=10\cmcube$ to $n=10^{3}\cmcube$ 
		(from the \textit{top} to \textit{bottom} panels).  
		We adopt the same colors as used in Figure 
		\ref{fig:211_x}, but different line types are used to distinguish between 
		the smaller or more compact binary ($a_{\rm s}$, dashed line) and the larger 
		or less compact binary ($a_{\rm l}$, solid line). The semimajor axis 
		estimated from the binding energy is shown along the upper x-axis. 
		 We provide the best fits in 
			Table \ref{tab:22_beta_SkewGaussian}. } 
	\label{fig:22_x}
\end{figure}

\begin{figure*}
	\centering
	\includegraphics[width=8.3cm]{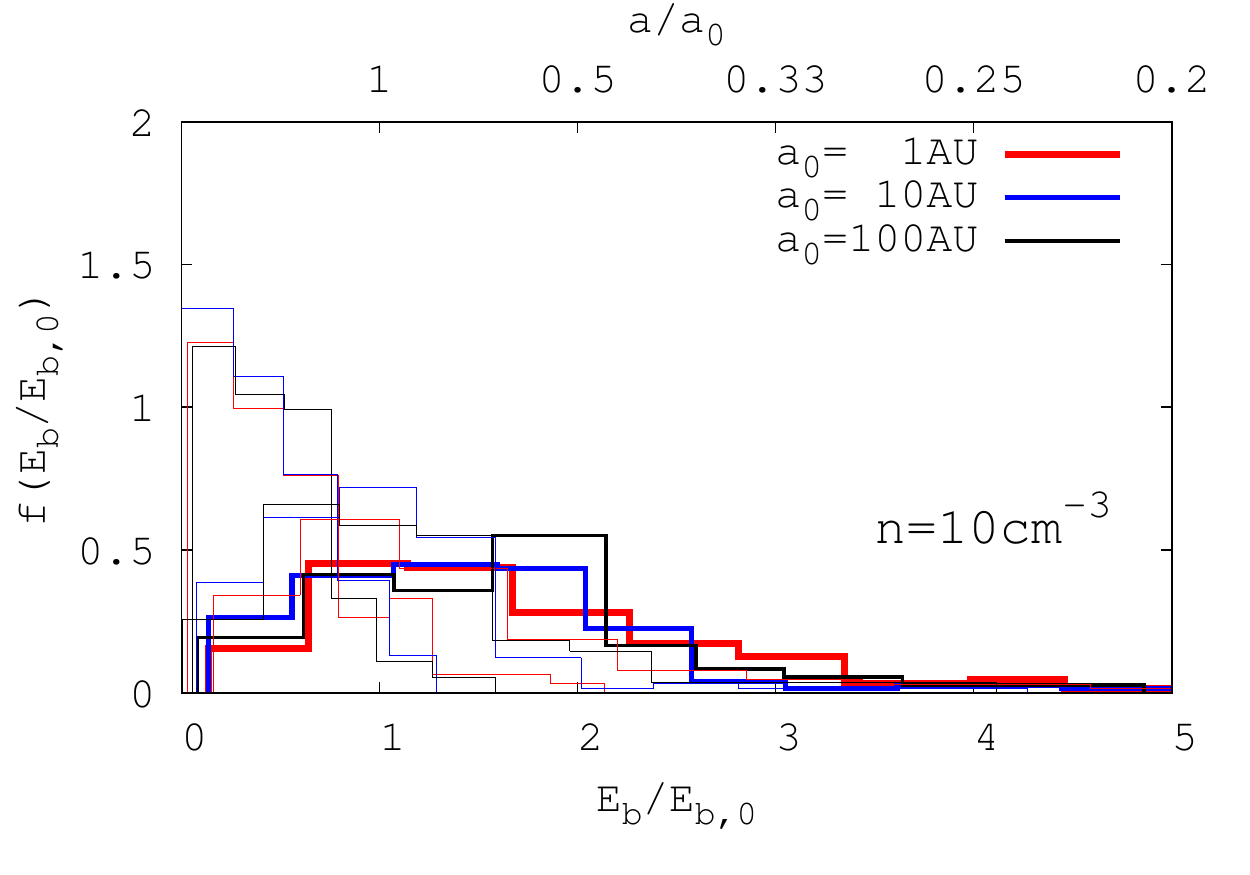}
	\includegraphics[width=8.3cm]{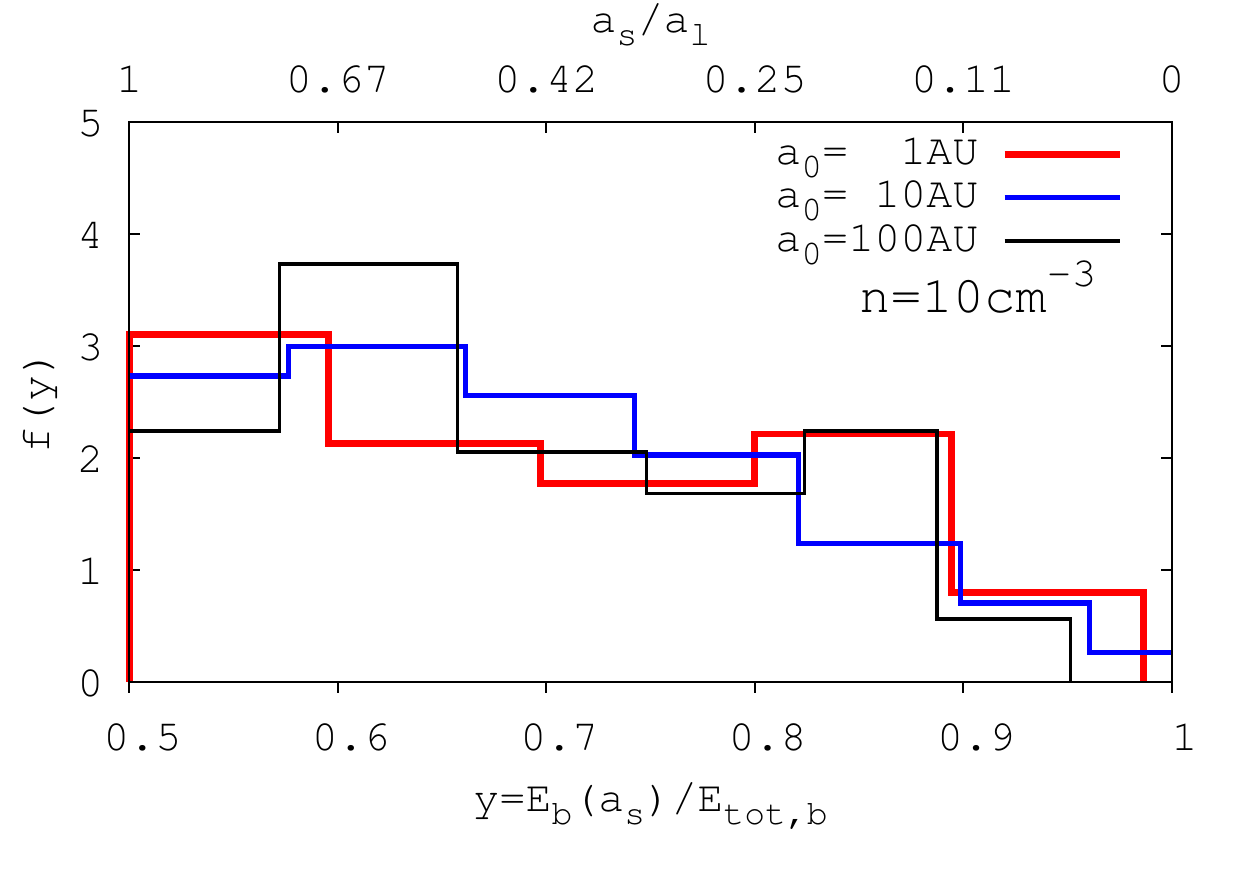}
	\includegraphics[width=8.3cm]{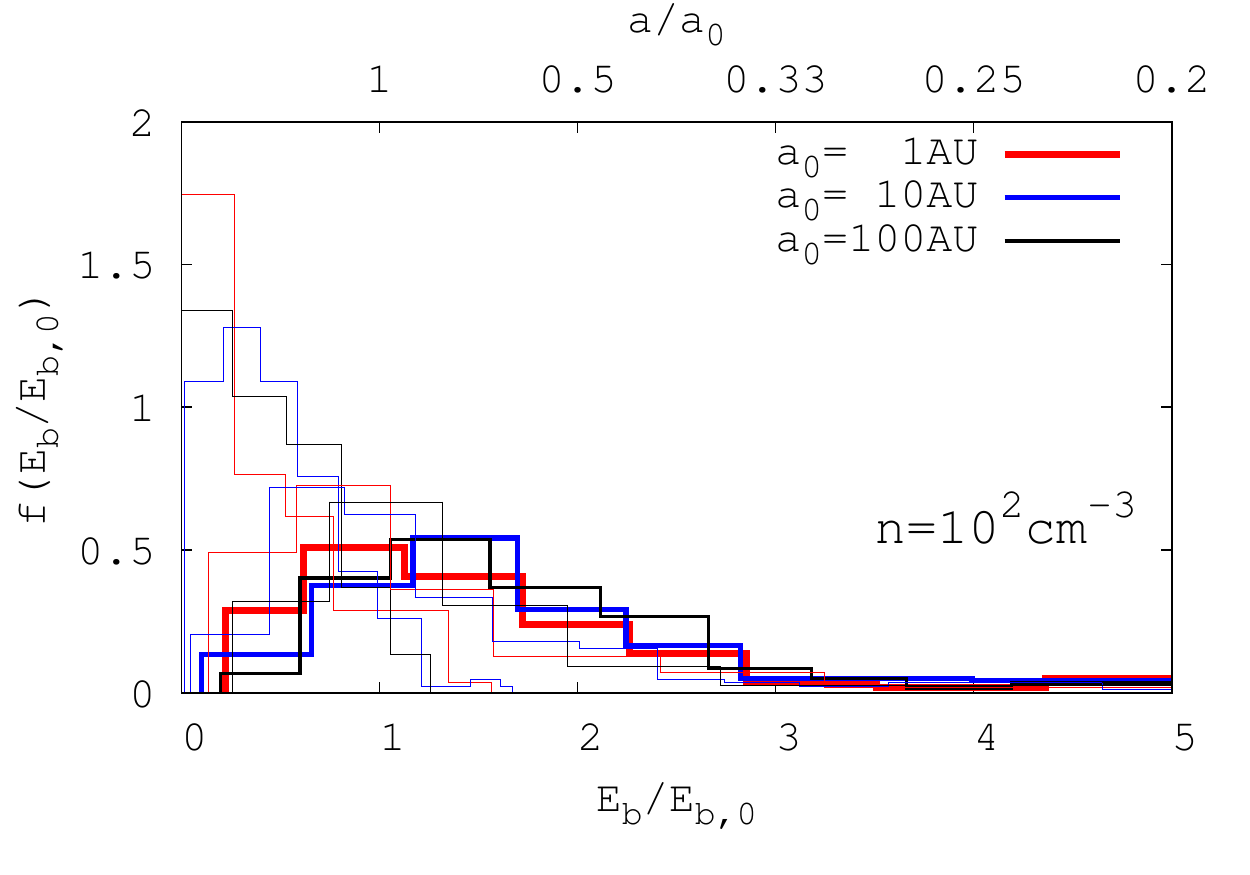}
	\includegraphics[width=8.3cm]{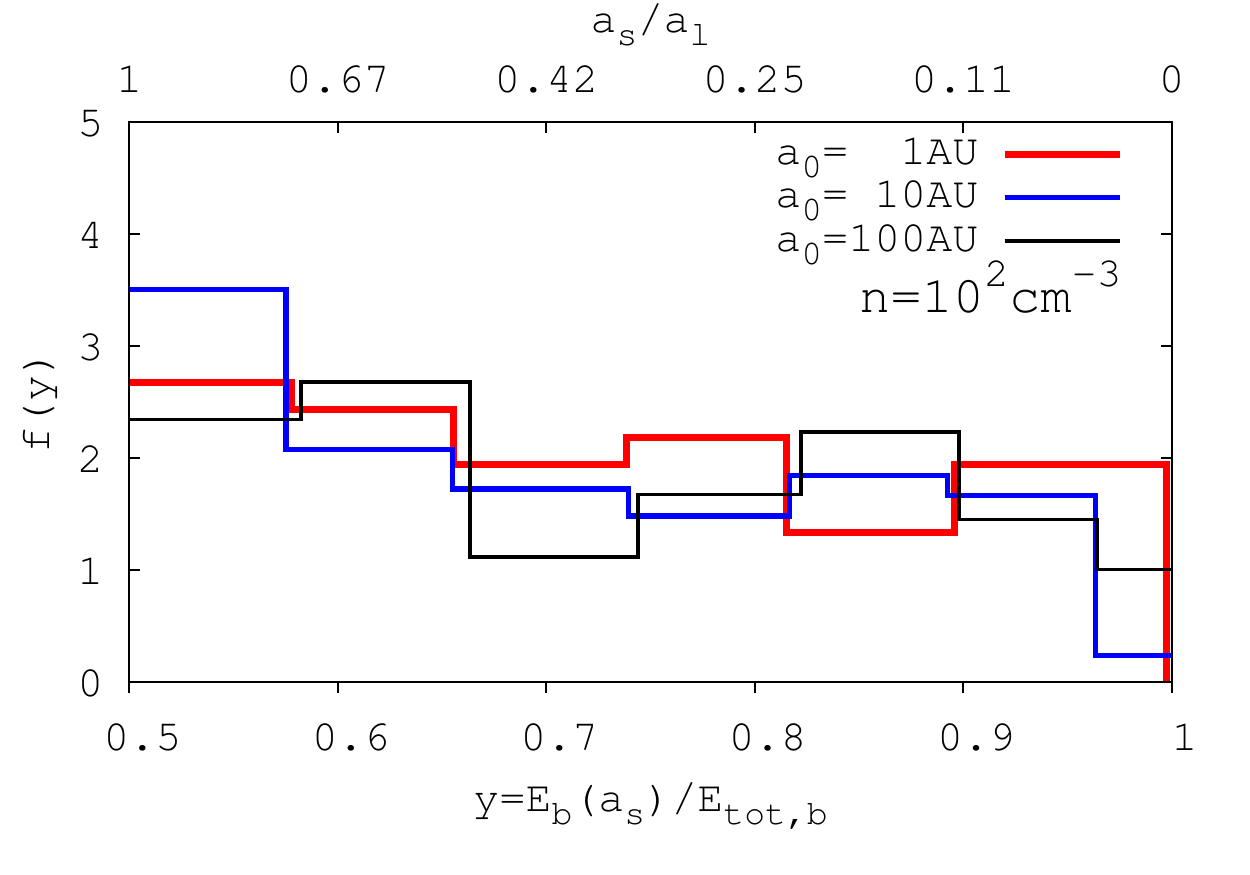}
	\includegraphics[width=8.3cm]{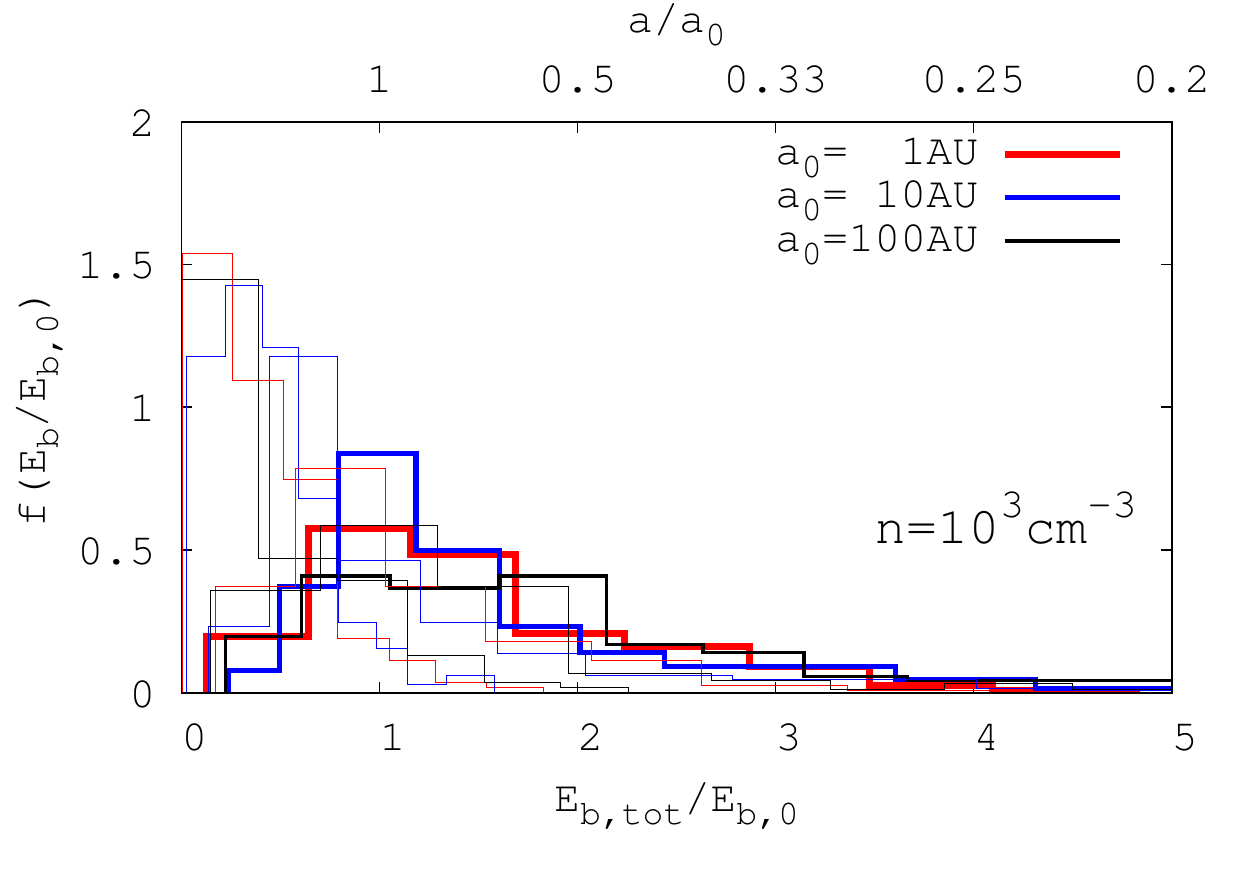}
	\includegraphics[width=8.3cm]{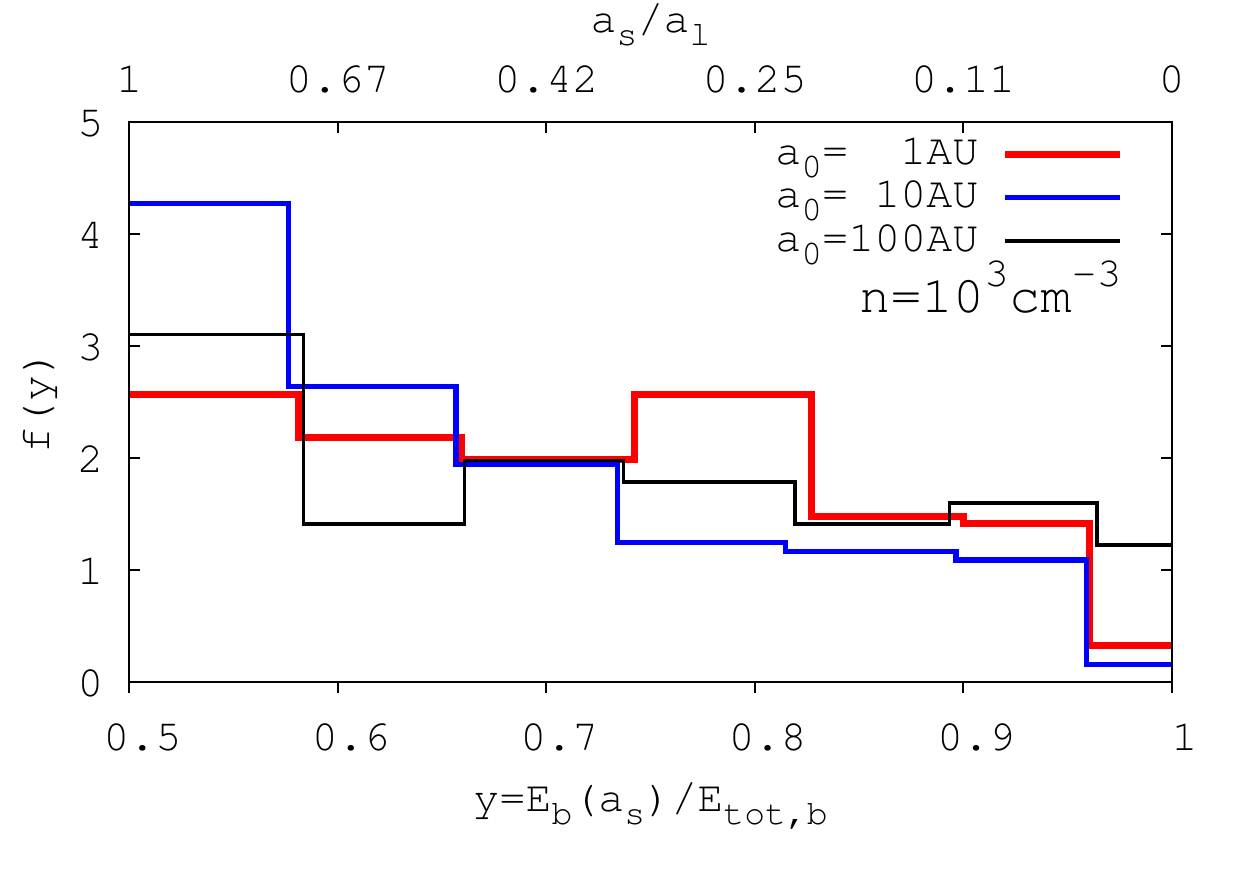}
	\caption{The final total binary binding energy distributions ($E_{\rm b, tot}$) 
		are shown with thick solid lines in the \textit{left} panel, and the fraction of this total $E_{\rm b, tot}$ 
                assigned to the more compact binary 
		is shown in the \textit{right panel} for number densities of $n=10\cmcube$ to $n=10^{3}\cmcube$ 
		(from the \textit{top} to \textit{bottom} panels).  
		In the \textit{left} panel, for comparison we over plot the distributions for each 
		individual binary shown in Figure \ref{fig:22_x} with thinner lines.
		 The semimajor axis estimated from each final 
                binding energy
		is indicated along the upper x-axis. In both panels, the same colors 
		are used as in Figure \ref{fig:211_x}. We provide the best fits for both distributions in 
			Table \ref{tab:22_beta_SkewGaussian} and \ref{tab:22_y_Gaussian}.}
	\label{fig:22_totx_y}
\end{figure*}

Figures~\ref{fig:211_x} and \ref{fig:211_e} show the distributions of final binary binding 
energies and eccentricities for each gas density and initial semimajor axis. In all four 
panels of Figure \ref{fig:211_x}, the red, blue and black solid lines correspond to, 
respectively, the initial semimajor axes $a_{0}=1\AU, 10\AU$ and $100\AU$. 
The distribution for $a_{0}=100\AU$ and $n=10^{7}\cmcube$ is 
	depicted by the green solid line in the \textit{top right} panel.
 The same colors are used for the initial semimajor axes as in Figure~\ref{fig:211_e}.

The final binary properties share several things in common independent 
of the gas density and initial semimajor axis. First, the final binary 
binding energy distribution has a peak at $E\simeq E_{\rm b,0}$, which corresponds to 
a semimajor axis for the final binary that is identical to the initial semimajor axis.  
That is, the binding energies of the two (identical) initial binaries 
determine that of the final binary. We find consistent results for the final relative 
velocities for the ejected single stars with respect to the final binary, in the sense 
that energy and momentum are conserved. 
The \textit{left} column in Figure \ref{fig:finalv} presents the final relative velocity
 distributions of single stars in the 2+1+1 case. 
 The same line colors and types are used in both columns as in Figure \ref{fig:211_x}. 
 The distributions have their peak at $v\simeq 0.6~v_{\rm cri}$. 
 Recall that, when $E_{0}\sim0$, the absolute value of 
 the binding energy is equal to the sum of the total kinetic energies 
 of the ejected single stars in the frame of reference of the final binary (with 
 reduced mass $2/3\Msol$).  We can also compare the final binary binding 
energy distribution for $E_{0}=0$  ($n=10^{2}\cmcube$ in the \textit{upper right} 
panel) with the ejection velocity distribution without a background potential 
in \citet{Leigh+16} ($k=1$ in their Figure 10), for the 2+1+1 outcome. 
Given that they find the same ejection velocities for both single stars in their 
ejection velocity distributions, corresponding to 
$KE(v_{\rm rel}=1)=|E_{b}|$ and $v_{\rm cri}(a_{0}=1\AU)=42\km \s^{-1}$,
we find good agreement with \citet{Leigh+16}. 
Second, all final binding energy
distributions extend up to $\sim12E/E_{b,0}$.  This implies that the semimajor
axis corresponding to the most compact binary that can possibly form via a head-on
collision at such high relative velocity is $\sim$ 1/12 
the initial semimajor axis. Third, given the low probability for $E<E_{b,0}$, 
the final binaries tend to be more compact than the initial binaries.  
But note that collisions with two binaries of unequal size
and non-zero impact parameter may yield different results.  
Finally, the eccentricity distributions in Figure \ref{fig:211_e} 
approximately follow a thermally-averaged density function
\citep{Heggie1975}, or:
\begin{equation}
 f(e)\sim 2e\,.
 \label{e_thermal}
 \end{equation} 
Interestingly, this is what is typically found for the binary eccentricity distributions 
formed during three-body scattering experiments \citep[i.e][]{Mikkola1994,Valtonen2006}. 
At least for the 2+1+1 outcome, four-body scatterings 
produce the same eccentricity distribution. With that said, we emphasize caution in 
concluding from this that a thermal distribution applies to all four-body scatterings.  
In particular, we will show a somewhat different eccentricity distribution for binaries 
formed during 2+2 outcomes in the following section. It is possible that 
the repeated scatterings (\textit{region~3}) are responsible for a slight departure for the $n=10^{5}$ 
case with $a_{0}=100\AU$ (black dashed line with cross mark) relative to the other cases. But, 
we cannot rule out statistical errors as being the primary effect here.

We find the best fits for the binding energy distributions assuming two different functional forms 
in Appendix \ref{appendix:fitting}. Motivated by \citet{Mikkola1983}, but 
with more free parameters ($a$, $b$, $c$ and $d$), we use the following expression:
\begin{equation}
f(x)\sim x(a+bx^{c})^{d}\,.
\end{equation}
For an alternative functional form, we use the skew normal distribution function, defined as the product of 
a normal distribution $\phi(x,\bar{x},\sigma_{x})$ and 
an accumulative normal distribution $\Phi(\xi,x,\bar{x},\sigma_{x})$ for a random variable $x$ 
with skewness $\xi$, or:
\begin{equation}
f(x)\sim \phi(x,\bar{x},\sigma_{x})\Phi(\xi,x,\bar{x},\sigma_{x})\,.
\end{equation}
All free parameters ($a$ to $d$ and $\bar{x}$, $\sigma_{x}$ and $\xi$) and 
the full expressions for the normal distribution $\phi(x,\bar{x},\sigma_{x})$ 
and accumulative normal distribution $\Phi(\xi,x,\bar{x},\sigma_{x})$ are summarized 
in Tables \ref{tab:211_Ebfitting} and \ref{tab:211_Eb_Gaussian}.

\subsubsection{The 2+2 outcome}

The 2+2 outcome has the second highest probability of occurring, at least until 
the 1+1+1+1 outcome dramatically escalates at $n=10^{3}\cmcube$.  Here,  
binary formation is suppressed due to the dissociation of any wide binary
by single stars when the stars are all trapped deep within the background potential.  
In the CM frame of the two final binaries, both binaries are ejected in opposite directions 
with the same velocity.  Only the final binary binding energies and/or 
eccentricities change. Since no single stars are formed during the 2+2 outcome, 
the total encounter energy can be decomposed into four components.  That is, letting 
$a_{\rm l}$ and $a_{\rm s}$ represent the semimajor axis of the 
larger and smaller binary, respectively, the total energy is:
\begin{align}
E&=E_{\rm b,tot}+KE_{\rm tot}\nonumber\\
&=E_{\rm b}(a_{\rm s})+E_{\rm b}(a_{\rm l})+KE(a_{\rm s})+KE(a_{\rm l})\,,
\label{eq:totalE_22}
\end{align}
where $E_{\rm b,tot}$ is the total binding energy (equal to the sum of the 
binding energies of the two binaries, or $E_{\rm b}(a_{\rm s})+E_{\rm b}(a_{\rm l})$) 
and $KE_{\rm tot}$ is the total kinetic energy (equal to the sum of the 
kinetic energies of the two binaries, or $KE(a_{\rm s})+KE(a_{\rm l})$). 

In Figure \ref{fig:22_x}, we compare the binding energies of the two binaries
formed during the 2+2 outcome, as a function of the gas density and the 
initial semimajor axis. The same color scheme is adopted 
as in Figure \ref{fig:211_x}, but two different line types are 
used to distinguish between the more (dashed line) and less (solid line) 
compact binary.  The distribution for $a_{\rm l}$ tends to be 
concentrated in the range $E_{b}/E_{b,0}<1$, which implies that the wider 
binary is less compact than the initial binary. The distribution for 
$a_{\rm s}$, on the other hand, peaks at $E_{b}/E_{b,0}\simeq1$.  Note that the 
overall shapes of the semimajor axis distributions are different than what we found 
for binaries formed during 2+1+1 outcomes (see Figure \ref{fig:211_x}). 

Importantly, as shown in Figure \ref{fig:22_x}, studying the binding energy 
distributions of each binary does not allow us to study the relationship 
between the two final binaries resulting 
from a particular interaction.
To correct this, in Figure \ref{fig:22_totx_y} we present the 
total binding energy ($E_{\rm b, tot}$) distribution of the two binaries combined (the thick 
solid line in the \textit{left} panel) and the fraction of the total binding energy assigned 
to the most compact binary ($y$ defined in equation \ref{eq:bindingenergyfraction} below) 
The \textit{right panels} show the results for number densities $n=10\cmcube$ to $n=10^{3}\cmcube$ 
(from \textit{top} to \textit{bottom}).  For comparison, we over plot in the \textit{left} panel 
the distributions for each individual binary (first shown in Figure \ref{fig:22_x}) 
with thinner lines. The semimajor axis associated with a given 
binding energy is indicated along the upper x-axis.

As seen in the \textit{left} panels, in which we combine the binding energy 
budgets from both binaries, 
$E_{\rm b,tot}$ reaches its maximum probability at $E_{b}/E_{b,0}\simeq 1 - 2$. 
This is typically the binding energy associated with each binary when they 
are ejected from each other at a relative velocity comparable to 
the escape velocity ($v_{\rm esc}$, which in our case is 
$v_{\rm esc} \geq v_{\rm cri}$). Interestingly, the high-energy tail in the 2+1+1 outcome 
distribution extends farther (i.e., out to $E_{b}/E_{b,0}\simeq 11 -12$) than 
in the 2+2 outcome distribution, which drops off rapidly at $E_{b}/E_{b,0}\simeq 5-6$. 
This is because the two single stars produced during the 2+1+1 outcome 
leave with more positive energy (in the form of kinetic energy), leaving 
a larger reservoir of negative energy for the final binary binding energy, relative 
to the 2+2 case.  Using a skew normal distribution, we obtain the 
best fitting parameters for the binding energy distribution
of each binary, as well as the total binding energy distribution.  These fit parameters 
are summarized in Table \ref{tab:22_beta_SkewGaussian}.

When the encounter is over, each binary takes some fraction of the total binary 
binding energy.  This is shown in the \textit{right} panels. Here, we introduce a variable $y$, 
defined as the fraction of the total binding energy assigned to the more compact binary:
\begin{align}
\label{eq:bindingenergyfraction}
y&=\frac{E_{\rm b}(a_{\rm s})}{E_{\rm tot,b}} \nonumber\\
&=\frac{a_{\rm l}}{a_{\rm s}+a_{\rm l}}=\frac{1}{a_{\rm s}/a_{\rm l}+1}\,.
\end{align}
The ratio $a_{\rm s}/a_{\rm l}$ corresponding to a given value of $y$ is 
marked along the upper x-axis. As $y$ approaches 0.5 (i.e., $a_{\rm s}/a_{\rm l}\simeq 1$), the two binaries share  
similar fractions of the total binary binding energy ($50$\% for each binary), 
and their semimajor axes are comparable.  However, if $y$ approaches unity, then 
one binary has most of the total binary binding energy; we are left 
with one compact binary and one wide binary.  
We find that the distributions decrease as $y$ increases, 
	implying that there are fewer outcomes in which one of the binaries 
	ends up with most of the total binding energy.
	In Table \ref{tab:22_y_Gaussian}, the best fitting values to these distributions 
	are reported, using the relation:
	\begin{equation}
	 f(y)=\alpha y +\beta\,.
	\end{equation}
Note that the slope $\alpha$ is always negative.

\begin{figure}
	\centering
	\includegraphics[width=8.3cm]{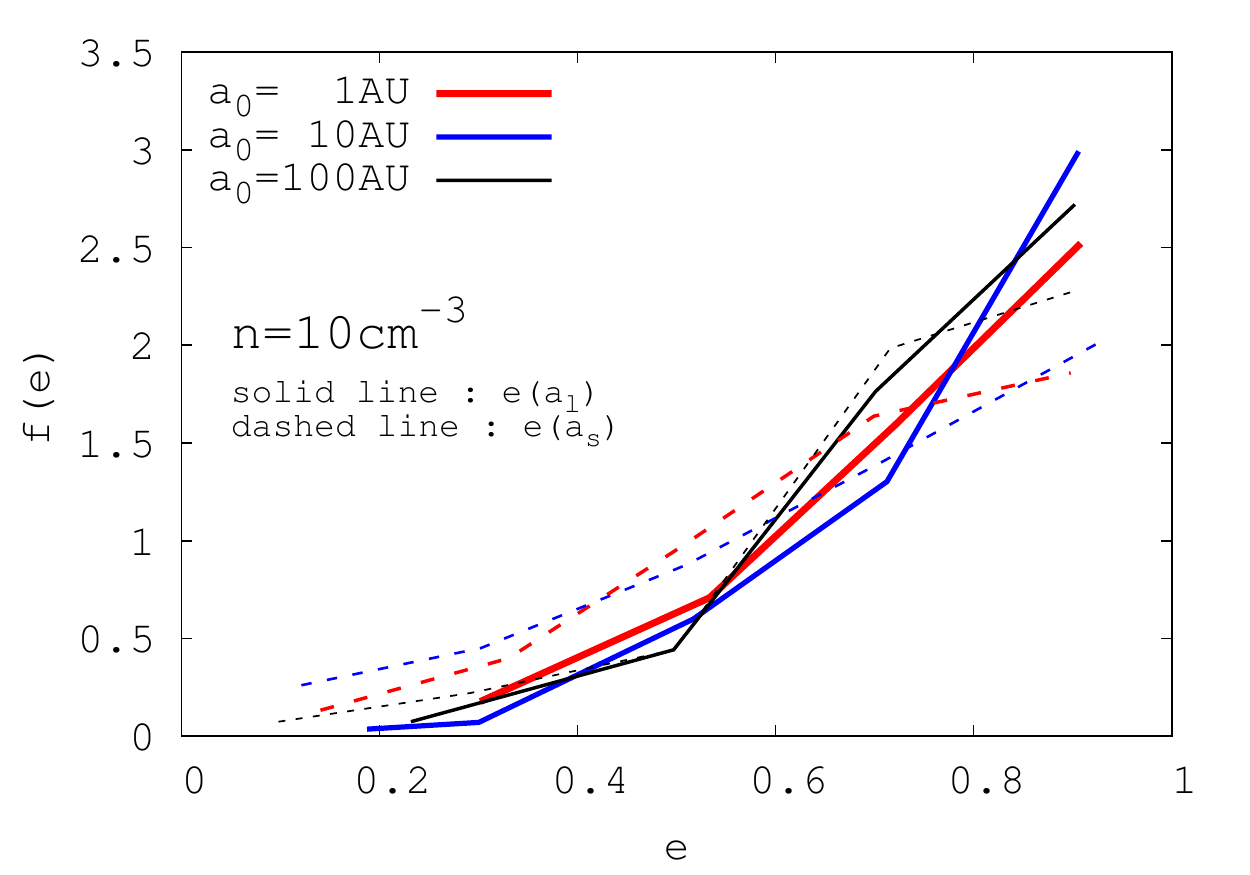}
	\includegraphics[width=8.3cm]{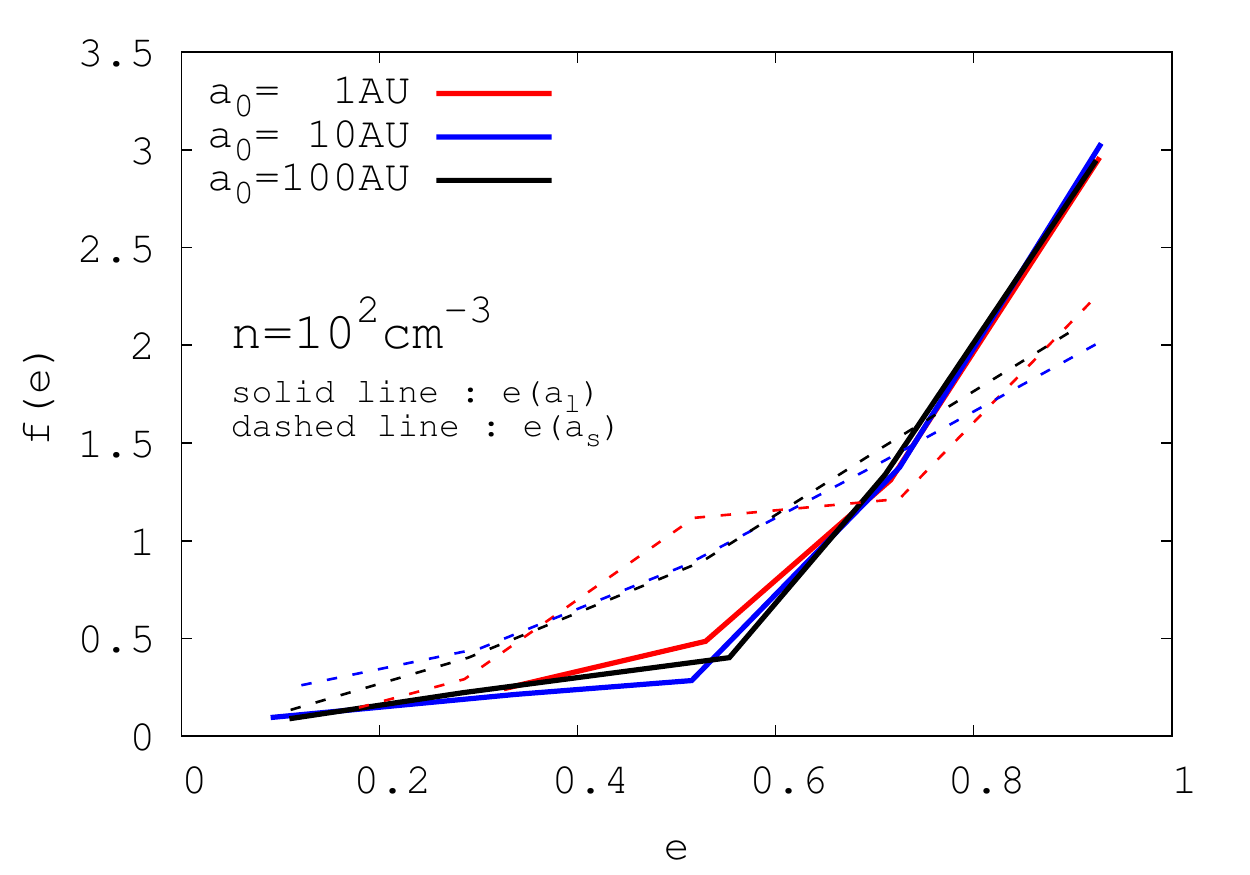}
	\includegraphics[width=8.3cm]{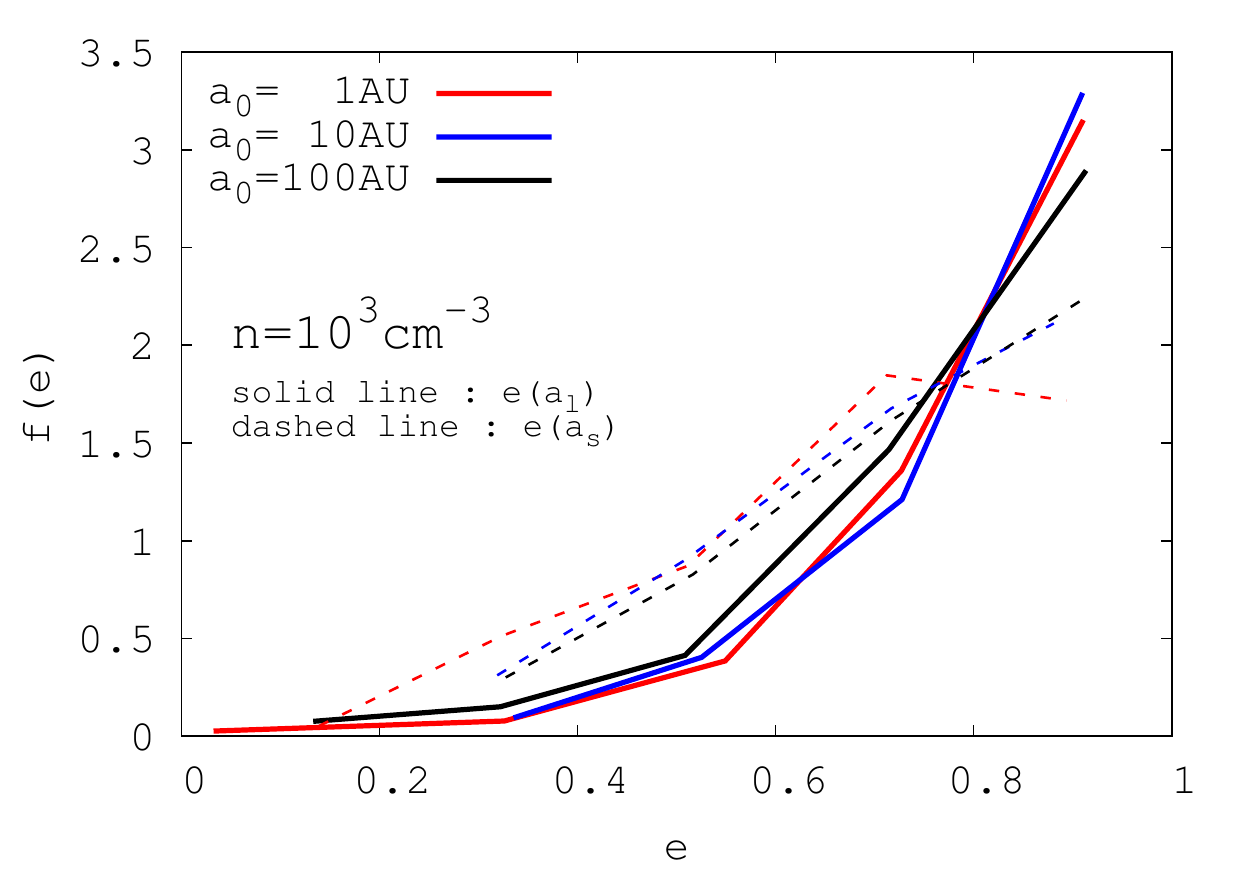}
	\caption{The eccentricity distributions are shown for binaries formed during 2+2 outcomes, 
		for number densities ranging from $n=10\cmcube$ to $n=10^{3}\cmcube$ 
		(from \textit{top} to \textit{bottom}) and different semimajor axes. 
		The solid and dashed lines correspond to the less and more 
		compact binary, respectively. The same
		color scheme is used as in Figure \ref{fig:211_x}. Note that all eccentricity 
		distributions follow the thermally averaged density function. 
		The best-fitting values of the parameter $\lambda$ are given in 
		Table \ref{tab:22_y_Gaussian}. }
	\label{fig:22_e}
\end{figure}

Finally, we show the eccentricity distributions for both binaries formed during 2+2 outcomes 
in Figure \ref{fig:22_e}. The same color scheme is used as before, but the solid 
and dashed lines correspond to the less ($a=a_{\rm l}$)
and more ($a=a_{\rm s}$) compact binary, respectively. The 
eccentricity distribution of the less compact binary does not follow the thermally averaged 
form (Equation \ref{e_thermal}). To characterize these distributions, we fit to 
the simulations the following general functional form with a free parameter $\lambda$:
\begin{equation}
f(e)\sim e^{\lambda}\,.
\end{equation}
The average value of $\lambda$ for $e (a_{\rm l})$ exceeds 2.3, whereas that 
for $e (a_{\rm s})$ is around $1.3 - 2.0$. Both of these values are larger than 1, 
which implies more eccentric binaries relative to the thermally averaged density function. 
We speculate that a value for $b$ larger than 1 arises due to 
conservation of angular momentum and energy,  
and the fact that a larger value 
for the semimajor axis requires a higher eccentricity to conserve angular momentum. 
The resulting values for $\lambda$ are given in Table \ref{tab:22_e_lambda}.

\section{Discussion and Summary}
\label{sec:discussionandsummary}

We have performed binary-binary scattering experiments in the presence 
of a background gravitational potential.  The background is taken to be an uniform
gas medium, for simplicity.  However, we emphasize that only the gravitational effects 
of the background potential are considered.  The potential remains static in time, 
and no gas damping (e.g., dynamical friction, accretion) can occur.  We explore a range 
of potential strengths corresponding to gas densities of $n=10\cmcube$ to 
$10^{5}\cmcube$ for three distinct values of the semimajor axis, namely $a=1\AU$, 
$10\AU$ and $100\AU$.  Additional simulations 
with other values of $a_{0}$ and $n$ were performed when needed.  

We find that the effect of the background potential on the 
scattering products depends on both the total energy of the system 
(whose zero-point is reset by the potential), as well as on its
relative strength compared to the potential generated by the single stars.

We summarize the effects of the background potential as follows:

\begin{enumerate}
	\item For any non-zero gas density $n$, the role of the 
	background potential is to \textit{reset the zero-point of the total system energy}.
	Accordingly, this affects the types and properties of the objects (single, binary and/or 
	triple stars) formed during the interactions as well as their outcome 
	probabilities. Since the contribution from the background potential to 
	the total system energy is always positive (for our assumption of an uniform 
	medium), it acts to decrease the zero-point of the total system energy.
	
	\item 
In order to isolate the various effects, and in particular the
relative role played by the initial binary semimajor axis $a_{0}$ for
a given gas density $n$, we have introduced three characteristic radii
($r_{\star}$, $r_{\rm return}$ and $r_{\rm bg}$), defined relative to
the system CM.  These radii define three distinct regions
(\textit{region~1}, \textit{region~2} and \textit{region~3}, from the
innermost to the outermost one), and play an important role relative to
$R$, the typical inter-particle separation during the scattering simulations in dense stellar
environments.
	
	If $R<r_{\star}$, the stellar interactions are confined to a region 
	where the stellar gravitational potential dominates over the contribution from 
	the background gas potential.  Hence, in this case, the effect of the 
	background potential on the coincident stellar dynamics is negligible. 
	The sets of experiments with ($n$, $a_{0}$)=($10-10^{2}\cmcube$, 
	$1-100\AU$) and ($10^{3}\cmcube$, $1-10\AU$) correspond 
	to this regime.  The outcome probabilities, as well as the types and properties 
	of the products, \textit{look reasonably
  	similar to the analogous experiments without a background potential}. 
	We compared our results with \citet{Leigh+16}. They performed 
	binary-binary scattering experiments exploring a range of values for the
	initial virial ratio. Our simulations with $n=10-10^{2}\cmcube$ 
	and $a_{0}=1\AU$ show good agreement with their results. 

	 If $r_{\star}<R\leq r_{\rm return}$, stars in this region begin to 
	 slow down as they feel the gravitational influence of the 
	 background potential (i.e., the enclosed gas mass starts to become comparable to 
	 the stellar mass).  We see this in the final velocity distributions in Figure 
	 \ref{fig:finalv}. For $n=10^{3}\cmcube$ and $a_{0}=100\AU$, 
	 the final velocity distributions are shifted to lower velocities. 
	 Even in this region, it is possible that stars become bound to the gas and 
	 return toward the system CM to 
	 go through additional scattering events. These scatterings act to 
	 ionize wide binaries and triples.
	Accordingly, this results in discrepancies in the 
	 outcome probabilities relative to the lower semimajor axis runs. Note, however, 
	 that our simulations do not show noticeable differences in the final binary 
	 properties. As we explain in detail in the next paragraph, the large differences appear with the emergence of very compact binaries due to numerous subsequent encounters. Hence, perhaps the number of encounters is insufficient for this to occur.  That is, since the ionization of the wide binaries or triples precedes the formation of the compact binaries, the outcome probabilities may start to change before the distributions do.
	 
	 If $ r_{\rm return}\leq R$, the background potential acts to ``filter out" 
	 certain outcomes at a given total energy. This is usually the case for 
	 high values of the gas density $n$. Due to the deep background 
	 potential well, stars remain bound to the gas and continue to interact 
	 until eventually they are ejected at sufficiently high velocity to become 
	 unbound and escape to infinity.  This requires strong, close gravitational interactions for 
	 the stars to gain these high ejection velocities.  Triples and wide binaries tend to 
	 form with low recoil velocities, such that they almost always remain bound to the background 
	 potential and return to undergo subsequent scattering events.  This process continues until 
	 eventually two single stars are ejected at high velocity, leaving behind a compact binary.   
	 \textit{Thus, in a deep background potential, stable triples and wide binaries are unlikely to form 
	 during binary-binary interactions.  Instead, the interactions tend to produce a very compact 
	 binary and two high-velocity single stars.}  
	 As a result, we see a shift in the binding energy distributions 
	 toward a higher energy range, and outcome probabilities around zero for the 2+2 and 3+1 outcomes. 
	 This is for the two sets of simulations with ($n$, $a_{0}$)=($10^{5}\cmcube$, $100\AU$) and ($10^{7}\cmcube$, $100\AU$).
  
	  \item The presence of the background potential can act to prolong the encounter 
     durations.  This is important in real astrophysical environments, and if one considers 
     finite-sized particles.  For example, the presence of the gas medium adds mass and increases 
     the gravitationally-focused cross-section for encounters with other stars in a star cluster.  This leads to 
     a non-negligible probability that the interactions will be interrupted \citep{GellerLeigh2015}.  
     This is especially true in astrophysical systems with 
     moderate stellar densities, such as the cores of open clusters.  Here, binary-binary encounters 
     should be frequently interrupted by other stars and binaries in the cluster \citep{GellerLeigh2015}.  
     If the encounters last longer, then the probability of a direct physical collision occurring also 
     increases \citep{LeighGeller2012,LeighGeller2015}.  This naively predicts that collisions and/or 
     mergers should happen more frequently during stellar dynamical interactions occurring in dense 
     gaseous environments.

\item 
In summary, we have discussed the stellar dynamics 
	in a continuous background potential 
	assuming an uniform gas medium.
We have shown from the changes in the outcome 
probabilities that the presence of the background potential resets
the zero-point of the total system energy. Furthermore, based on the three characteristic distances 
($r_{\rm star}$, $r_{\rm return}$ and 
$r_{\rm gas}$) which quantify the volume corresponding to each of the three regions, 
we find a relation between $a_{0}$ and $n$ (equation \ref{eq:threeregions}). 
This relation, in turn, allows us to identify when
 the background potential makes significant impacts on the dynamics between stars, 
 that is, when the presence of the gas causes deviations from what should be 
 expected for the 4-body dynamics in isolation.

\end{enumerate}

\vspace{0.5cm}

\section*{Acknowledgements}

We are grateful to the referee for constructive and meaningful comments. 
Results in this paper were obtained using the
high-performance LIred computing system at the Institute for Advanced
Computational Science at Stony Brook University, which was obtained
through the Empire State Development grant NYS \#28451.








\appendix

\section{Fitting formulae for distribution functions}
\label{appendix:fitting}

\begin{table*}
	\centering
	\setlength\extrarowheight{4pt}
	\begin{tabulary}{0.7\linewidth}{ c| c c c| ccc|ccc }
		\hline
		\hline
		& \multicolumn{3}{c|}{	$n=10\cmcube$} & \multicolumn{3}{c|}{	$n=10^{2}\cmcube$}& \multicolumn{3}{c}{	$n=10^{2}\cmcube$}\\
		\hline
		$	a_{0}$	   &  $~~1\AU$& $~10\AU$    &  $100\AU$   &  $~~1\AU$& $~10\AU$    &  $100\AU$   &  $~~1\AU$& $~10\AU$    &  $100\AU$       \\   
		\hline
		\hline
		$a$ 	 & 1.0743        &0.8307      &  0.9667     &0.9351    &0.9344           &  0.6465    &      0.0659  & 0.1035       &  $1.6222\times10^{-4}$     \\
		
		$b$      & 0.0235 	     & 0.0851    & 0.0399    &0.0635  &  0.0353        &  0.0583     &    0.01212     &   $ 8.0138\times10^{-3}$      &  $6.0042\times10^{-6}$       \\
		
		$c$      & 1.4334	      & 2.6455        & 1.6667        &1.8940      &  4.1905             &   1.5036      &  4.8179     &  1.7381      & 4.1526    \\
		
		$d$	     &  -15.1900     &   -1.5990    &  -6.9951     &-3.2957    &  -0.9147        &  -4.7687     &   -0.6612     &  -3.9653         &   -0.9783       \\
		\hline
		\hline
	\end{tabulary}
	\caption{Best fits for the binding energy distributions for the 2+1+1 outcome with $f(x)\sim x(a+bx^{c})^{d}$.}
	\label{tab:211_Ebfitting}
\end{table*}

\begin{table*}
	\centering
	\setlength\extrarowheight{4pt}
	\begin{tabulary}{0.7\linewidth}{ c| c c c| ccc|ccc }
		\hline
		\hline
		$x=E_{\rm b}/E_{\rm b ,0}$& \multicolumn{3}{c|}{	$n=10\cmcube$} & \multicolumn{3}{c|}{	$n=10^{2}\cmcube$}& \multicolumn{3}{c}{	$n=10^{2}\cmcube$}\\
		\hline
		$	a_{0}$	   &  $~~1\AU$& $~10\AU$    &  $100\AU$   &  $~~1\AU$& $~10\AU$    &  $100\AU$   &  $~~1\AU$& $~10\AU$    &  $100\AU$       \\   
		\hline
		\hline
		$\bar{x}$ 	 & 2.5749        &2.2940    & 2.4367    &2.6379     &2.3373         &  2.2748     & 1.7118        & 2.3937          &  2.3000       \\
		
		$\sigma_{x}$      & 2.0790	     &  1.7851  & 1.7848  &1.9925 &  1.4324       &    2.2126     &  1.3159        &   1.9919      &  1.3764   \\
		\hline
		\hline
	\end{tabulary}
	\caption{Best fits for the binding energy distributions for the 2+1+1 outcome with the skew normal distribution $f(x)\sim \phi(x,\bar{x},\sigma_{x})\Phi(\xi,x,\bar{x},\sigma_{x})$ where $\phi(x,\bar{x},\sigma_{x})=\frac{1}{\sqrt{2\pi \sigma_{x}^{2}}}e^{-(x-\bar{x})^{2}/\sigma_{x}^{2}}$ and $\Phi(\xi,x,\bar{x},\sigma_{x})=\int_{\infty}^{\xi (x-\bar{x})/\sigma_{x}}e^{-\frac{t^{2}}{2}}dt=\frac{1}{2}\Big[1+{\rm erf} \Big(\xi\frac{x-\bar{x}}{\sqrt{2}\sigma_{x}}\Big)\Big]$. We take the skewness $\alpha=1.0$.  	}
	\label{tab:211_Eb_Gaussian}
\end{table*}

We provide the best fits for the binding energy and eccentricity distributions shown in the text for the 2+1+1 and 2+2 outcomes.

\subsection{2+1+1 case}
For the binding energy, we fit two functional forms. First, motivated by 
\citet{Mikkola1983} but 
with more free parameters ($a$, $b$, $c$ and $d$), we use the following:
\begin{equation}
f(x)\sim x(a+bx^{c})^{d}\,.
\end{equation}
For the second fit, we use the skew normal distribution function, defined as the product of a 
normal distribution $\phi(x,\bar{x},\sigma_{x})$ and an 
accumulative normal distribution $\Phi(\xi,x,\bar{x},\sigma_{x})$ for a random variable $x$ with skewness $\xi$,
\begin{equation}
f(x)\sim \phi(x,\bar{x},\sigma_{x})\Phi(\xi,x,\bar{x},\sigma_{x})\,.
\end{equation}
All free parameters ($a$ to $d$ and $\bar{x}$, $\sigma_{x}$ and $\xi$) and 
the full expressions for the normal distribution $\phi(x,\bar{x},\sigma_{x})$ 
and accumulative normal distribution $\Phi(\xi,x,\bar{x},\sigma_{x})$ are summarized 
in Table \ref{tab:211_Ebfitting} and \ref{tab:211_Eb_Gaussian}. We take the skewness $\xi=1$.
The eccentricities for the 2+1+1 outcome follow a thermally averaged density distribution, $f(e)\sim e$ (equation \ref{e_thermal}).

\newpage
\subsection{2+2 case}

\begin{table*}
	\centering
	\setlength\extrarowheight{4pt}
	\begin{tabulary}{0.7\linewidth}{ m{1.5cm} c| m{1.23cm} m{1.5cm} m{1.23cm} | m{1.23cm} m{1.25cm} m{1.23cm} |m{1.23cm} m{1.5cm} m{1.23cm} }
		\hline
		\hline
		\multicolumn{2}{c|}{\multirow{2}{*}{$x=E_{\rm b}/E_{\rm b ,0}$}}& \multicolumn{3}{c|}{	$a_{0}=1\AU$} & \multicolumn{3}{c|}{$a_{0}=10\AU$}& \multicolumn{3}{c}{	$a_{0}=100\AU$}\\
		\cline{3-11}
		&   & $a_{\rm s}$    & $a_{\rm l}$&  total    & $a_{\rm s}$    & $a_{\rm l}$&  total   & $a_{\rm s}$    & $a_{\rm l}$&  total        \\   
		\hline
		\hline
		\multirow{2}{*}{$n=10~\cmcube$}&	$\bar{x}$ 	 &  $\text{\scriptsize 1.169} _{\pm\text{\tiny 0.037 }  }$     &   $\text{\scriptsize 0.204}_{\pm\tiny 0.192  }$   &$\text{\scriptsize 1.634} _{\pm\text{\tiny 0.087}  }$         & $\text{\scriptsize 1.065} _{\pm\text{\tiny 0.028}  }$   &    $\text{\scriptsize 0.277}_{\pm\text{\tiny 0.048} }$          &   $\text{\scriptsize  1.540} _{\pm\text{\tiny 0.051}  }$    & $\text{\scriptsize  1.130}_{\pm\text{\tiny 0.060}  }$          &    $\text{\scriptsize  0.427} _{\pm\text{\tiny 0.082}}$      &     $\text{\scriptsize 1.738}_{\pm\text{\tiny 0.116}  }$      \\
		
		&	$\sigma_{x}$         &   $\text{\scriptsize 0.640} _{\pm\text{\tiny 0.048}  }$    &     $\text{\scriptsize 0.642}_{\pm\text{\tiny 0.133}}$   &  $\text{\scriptsize 0.892} _{\pm\text{\tiny 0.103}  }$    &  $\text{\scriptsize 0.570} _{\pm\text{\tiny 0.033}  }$ &      $\text{\scriptsize 0.539}_{\pm\text{\tiny  0.0451}} $        & $\text{\scriptsize 0.862} _{\pm\text{\tiny 0.064}  }$    &   $\text{\scriptsize  0.628} _{\pm\text{\tiny  0.071 }  }$         &     $\text{\scriptsize 0.425}_{\pm\text{\tiny 0.107} }$         &  $\text{\scriptsize  0.832} _{\pm\text{\tiny 0.138}  }$   \\
		\hline
		\multirow{2}{*}{$n=10^{2}\cmcube$}&	$\bar{x}$ 	 &  {\scriptsize0.937}$_{\pm}${\tiny 0.030}        & {\scriptsize-5.022}$_{\pm}${\tiny 30.82}     & {\scriptsize1.368}$_{\pm}${\tiny 0.052}     & {\scriptsize0.994}$_{\pm}${\tiny 0.045}  &   {\scriptsize0.441}$_{\pm}${\tiny 0.013}      &  {\scriptsize1.632}$_{\pm}${\tiny 0.041}     &   {\scriptsize1.229}$_{\pm}${\tiny 0.030}             &   {\scriptsize0.297}$_{\pm}${\tiny 0.136}        & {\scriptsize1.699}$_{\pm}${\tiny 0.086}   \\
		
		&	$\sigma_{x}$     &   {\scriptsize0.527}$_{\pm}${\tiny 0.039}      & {\scriptsize1.561}$_{\pm}${\tiny 4.294}   &  {\scriptsize0.831}$_{\pm}${\tiny 0.076}  & {\scriptsize0.472}$_{\pm}${\tiny 0.049}  &       {\scriptsize0.428}$_{\pm}${\tiny 0.017}   &  {\scriptsize0.693}$_{\pm}${\tiny 0.044}   &   {\scriptsize0.517}$_{\pm}${\tiny 0.032}     &      {\scriptsize0.558}$_{\pm}${\tiny 0.133}         &   {\scriptsize0.773}$_{\pm}${\tiny 0.089} \\
		\hline
		\multirow{2}{*}{$n=10^{3}\cmcube$}&	$\bar{x}$ 	&   {\scriptsize0.984}$_{\pm}${\tiny 0.056}        &   {\scriptsize0.177}$_{\pm}${\tiny 0.106}  &   {\scriptsize1.355}$_{\pm}${\tiny 0.082}   &  {\scriptsize0.750}$_{\pm}${\tiny 0.029}     &    {\scriptsize0.420}$_{\pm}${\tiny 0.009}     &   {\scriptsize1.210}$_{\pm}${\tiny 0.044}    &     {\scriptsize1.228}$_{\pm}${\tiny 0.028}       &   {\scriptsize-4.791}$_{\pm}${\tiny 23.9}         &     {\scriptsize 1.698}$_{\pm}${\tiny 0.084}   \\
		
		&	$\sigma_{x}$     & {\scriptsize0.450}$_{\pm}${\tiny0.063}  &  {\scriptsize0.504}$_{\pm}${\tiny 0.077}     &  {\scriptsize0.641}$_{\pm}${\tiny 0.092}     & {\scriptsize0.233}$_{\pm}${\tiny 0.028} &  {\scriptsize0.334}$_{\pm}${\tiny 0.012}        &   {\scriptsize0.402}$_{\pm}${\tiny 0.045}    &      {\scriptsize0.657}$_{\pm}${\tiny 0.034}    & {\scriptsize1.610}$_{\pm}${\tiny 3.523}          &{\scriptsize0.992}$_{\pm}${\tiny 0.111}  \\
		\hline
		\hline
	\end{tabulary}
	\caption{Best fits for the binding energy distributions for the 2+2 outcome with the skew normal distribution $f(x)\sim \phi(x,\bar{x},\sigma_{x})\Phi(\xi,x,\bar{x},\sigma_{x})$ where $\phi(x,\bar{x},\sigma_{x})=\frac{1}{\sqrt{2\pi \sigma_{x}^{2}}}e^{-(x-\bar{x})^{2}/\sigma_{x}^{2}}$ and $\Phi(\xi,x,\bar{x},\sigma_{x})=\int_{\infty}^{\xi (x-\bar{x})/\sigma_{x}}e^{-\frac{t^{2}}{2}}dt=\frac{1}{2}\Big[1+{\rm erf} \Big(\xi\frac{x-\bar{x}}{\sqrt{2}\sigma_{x}}\Big)\Big]$. We take the skewness $\alpha=1.0$.  }
	\label{tab:22_beta_SkewGaussian}
\end{table*}

\begin{table*}
	\centering
	\setlength\extrarowheight{3pt}
	\begin{tabulary}{0.7\linewidth}{ c| m{1.5cm} m{1.5cm}m{1.5cm}| m{1.5cm}m{1.5cm}m{1.5cm}|m{1.5cm}m{1.5cm}m{1.5cm} }
		\hline
		\hline
		$y$ & \multicolumn{3}{c|}{	$n=10\cmcube$} & \multicolumn{3}{c|}{	$n=10^{2}\cmcube$}& \multicolumn{3}{c}{	$n=10^{2}\cmcube$}\\
		\hline
		$	a_{0}$	   &  $~~1\AU$& $~10\AU$    &  $100\AU$   &  $~~1\AU$& $~10\AU$    &  $100\AU$   &  $~~1\AU$& $~10\AU$    &  $100\AU$       \\   
		\hline
		\hline
		slope $\alpha$	   &   {\scriptsize-4.516}$_{\pm}${\tiny 1.606}        &  {\scriptsize-6.132}$_{\pm}${\tiny 0.834}    &   {\scriptsize-4.775}$_{\pm}${\tiny 2.519}  &  {\scriptsize-2.253}$_{\pm}${\tiny 1.039 }     &  {\scriptsize-4.804}$_{\pm}${\tiny 1.406}          &   {\scriptsize-2.477}$_{\pm}${\tiny 1.353}    &   {\scriptsize-3.963}$_{\pm}${\tiny 1.224}         &   {\scriptsize-7.568}$_{\pm}${\tiny 1.284}          &    {\scriptsize-2.780}$_{\pm}${\tiny 1.169}      \\
		
		intercept  $\beta$     &    {\scriptsize5.370}$_{\pm}${\tiny 1.219}     &   {\scriptsize6.535}$_{\pm}${\tiny 0.659}   &   {\scriptsize5.597}$_{\pm}${\tiny 1.884 }   &  {\scriptsize3.745}$_{\pm}${\tiny 0.780}  &    {\scriptsize5.495}$_{\pm}${\tiny 1.107}    &    {\scriptsize3.711}$_{\pm}${\tiny 1.071}     &    {\scriptsize4.862}$_{\pm}${\tiny 0.968}       &      {\scriptsize7.621}$_{\pm}${\tiny 1.009 }    &    {\scriptsize3.941}$_{\pm}${\tiny 0.924 }  \\
		\hline
		\hline
	\end{tabulary}
	\caption{Best fits for $y$ for the 2+2 outcome with $f(y)=\alpha y +\beta$.	}
	\label{tab:22_y_Gaussian}
\end{table*}

For the binding energies of each binary (wide and compact) and the total 
binding energy ($E_{\rm t, tot}$), we use the skew normal distribution function. 
This is the same as used for the binding energy distribution for the 2+1+1 outcome. The average values 
and the standard deviations of our best fits are summarized
 in Table \ref{tab:22_beta_SkewGaussian}. 
We also  take the skewness $\xi=1$.

Next, we define the variable $y$ as the fraction of the binding energy assigned 
to the more compact binary. Each binary takes some fraction of the total binding energy,
\begin{align}
y&=\frac{E_{\rm b}(a_{\rm s})}{E_{\rm tot,b}} \nonumber \\
&=\frac{a_{\rm l}}{a_{\rm s}+a_{\rm l}}=\frac{1}{a_{\rm s}/a_{\rm l}+1}\,.
\end{align}
For the best fit, we use the following formula,
	\begin{equation}
	f(y)=\alpha y +\beta\,.
	\end{equation}
	Two parameters $\alpha$ (slope) and $\beta$	(intercept) in Table \ref{tab:22_y_Gaussian}.

Finally, for the eccentricities of each binary (wide and compact), we fit to 
the simulations the following general functional form with a free parameter $\lambda$:
\begin{equation}
f(e)\sim e^{\lambda}\,.
\end{equation}
the values for $\lambda$ are given in Table \ref{tab:22_e_lambda}. Maximum 
and minimum values for $e$ are written below each value of $\lambda$.
\begin{table*}
	\centering
	\setlength\extrarowheight{4pt}
	\begin{tabulary}{0.7\linewidth}{ c| c c |c c| c c }
		\hline
		\hline
		\multirow{3}{*}{ $n~\backslash~a_{0}$}    &\multicolumn{2}{c|}{$1\AU$} & \multicolumn{2}{c|}{$10\AU$}  & \multicolumn{2}{c}{$100\AU$ }\\ \cline{2-7}
		&    $a_{\rm s}$     &  $a_{\rm l}$  	&    $a_{\rm s}$     &  $a_{\rm l}$  	&    $a_{\rm s}$     &  $a_{\rm l}$  \\   
		& $^{[	e_{\rm min}	-	e_{\rm max}		]} $ & $^{[	e_{\rm min}	-	e_{\rm max}		]} $& $^{[	e_{\rm min}	-	e_{\rm max}		]} $& $^{[	e_{\rm min}	-	e_{\rm max}		]} $& $^{[	e_{\rm min}	-	e_{\rm max}		]} $& $^{[	e_{\rm min}	-	e_{\rm max}		]} $\\
		\hline
		\hline
		\multirow{2}{*}{$10^{}\cmcube$}	 &  1.2990$\pm${\scriptsize 0.2193}/& 2.2729$\pm${\scriptsize 	0.1506}  &  1.2965$\pm${\scriptsize 0.0994} & 3.1814$\pm${\scriptsize 0.1878} &  1.9646$\pm${\scriptsize 0.5764 }  & 2.4125$\pm${\scriptsize 0.4754} \\
		&$^{[	0.0789	-	0.9947	]} $                    & $^{[	0.2838	-	0.9970	]}$   &$^{[	0.0634	-	0.9945	]}$& $^{[	0.1874	-	0.9985	]}$ & 
		$^{[	0.0981	-	0.9973	]}$ &$^{[	0.2318	-	0.9977	]}$\\
		\multirow{2}{*}{$10^{2}\cmcube$}	&1.5537$\pm${\scriptsize 0.3234}  &  3.0974$\pm${\scriptsize 0.2209}  & 1.2965$\pm${\scriptsize 0.0994 }  &  3.3910$\pm${\scriptsize 0.3073}   & 1.4791$\pm${\scriptsize 0.0571}  &  3.2876$\pm${\scriptsize 0.3606} \\
		& $^{[	0.1482	-	0.9968	]}$ & $^{[	0.2456	-	0.9979	]}$ & $^{[	0.0000	-	0.9996	]}$ & $^{[	0.0000	-	0.9997	]}$ &$^{[	0.0503	-	0.9969	]}$ &$^{[	0.0923	-	0.9993	]}$\\
		\multirow{2}{*}{$10^{3}\cmcube$}  &  1.2338$\pm${\scriptsize 0.3610}&  3.88354$\pm${\scriptsize 0.1297} & 1.6341$\pm${\scriptsize 0.2023 } &  4.2376$\pm${\scriptsize 0.2541} & 1.7470$\pm${\scriptsize 0.1731} &  2.9867$\pm${\scriptsize 0.1820}\\
		& $^{[	0.1350	-	0.9876	]}$ &$^{[	0.0324	-	0.9998	]}$ &$^{[	0.2400	-	0.9941	]}$ & $^{[	0.3163	-	0.9988	]}$ & $^{[	0.2743	-	0.9991	]}$ & $^{[	0.0693	-	0.9997	]}$\\
		\hline
		\hline
	\end{tabulary}
	\caption{Best fits for the eccentricities for each binary for the 2+2 outcome with $f(y)\sim e^{\lambda}$. Maximum and minimum values for $e$ are written below each value of $\lambda$. }
	\label{tab:22_e_lambda}
\end{table*}


\bsp	
\label{lastpage}
\end{document}